# Resolving the Disc-Halo Degeneracy I: A Look at NGC 628

S. Aniyan[1,2]⋆, K. C. Freeman[1], M. Arnaboldi[2], O. E. Gerhard[3], L. Coccato[2], M. Fabricius[3], K. Kuijken[4], M. Merrifield[5] & A. A. Ponomareva[1,6]

[1]*Research School of Astronomy & Astrophysics, Australian National University, Canberra, ACT 2611, Australia*
[2]*European Southern Observatory, Karl-Schwarzschild-Strasse 2, D-85748 Garching, Germany*
[3]*Max-Planck-Institut für Extraterrestrische Physik, Giessenbachstrasse, 85741 Garching, Germany*
[4]*Leiden Observatory, Leiden University, Niels Bohrweg 2, NL-2333 CA Leiden, the Netherlands*
[5]*School of Physics and Astronomy, University of Nottingham, University Park, Nottingham, NG7 2RD, UK*
[6]*Kapteyn Astronomical Institute, University of Groningen, Postbus 800, NL-9700 AV Groningen, The Netherlands*

1 February 2018

**ABSTRACT**
The decomposition of the rotation curve of galaxies into contribution from the disc and dark halo remains uncertain and depends on the adopted mass to light ratio ($M/L$) of the disc. Given the vertical velocity dispersion of stars and disc scale height, the disc surface mass density and hence the $M/L$ can be estimated. We address a conceptual problem with previous measurements of the scale height and dispersion. When using this method, the dispersion and scale height must refer to the same population of stars. The scale height is obtained from near-IR studies of edge-on galaxies and is weighted towards older kinematically hotter stars, whereas the dispersion obtained from integrated light in the optical bands includes stars of all ages. We aim to extract the dispersion for the hotter stars, so that it can then be used with the correct scale height to obtain the disc surface mass density. We use a sample of planetary nebulae (PNe) as dynamical tracers in the face-on galaxy NGC 628. We extract two different dispersions from its velocity histogram – representing the older and younger PNe. We also present complementary stellar absorption spectra in the inner regions of this galaxy and use a direct pixel fitting technique to extract the two components. Our analysis concludes that previous studies, which do not take account of the young disc, underestimate the disc surface mass density by a factor of ∼2. This is sufficient to make a maximal disc for NGC 628 appear like a submaximal disc.

## 1 INTRODUCTION

The 21 cm rotation curve of galaxies flatten at large radii, indicating the presence of dark matter in these galaxies. The rotation curves can be decomposed into contributions from the stellar and gas discs, plus the dark halo, and in principle allow us to estimate the parameters of the dark halo. The decomposition of these rotation curves into contributions from the disc and the dark halo depends strongly, however, on the adopted mass-to-light ratio ($M/L$) of the stellar disc (Van Albada et al. 1985). Choosing different $M/L$ can result in a maximal disc or a submaximal disc, with very different dark halo contributions, both of which can fit the observed rotation curves equally well. Thus, the $M/L$ is critical to obtain the parameters of the dark haloes of disc galaxies, such as their scale densities and scale lengths.

These halo parameters are cosmologically significant, because the densities and scale radii of dark haloes follow well-defined scaling laws and can therefore be used to measure the redshift of assembly of haloes of different masses (Macciò et al. 2013; Kormendy & Freeman 2016).

Several techniques have been used to break the disc-halo degeneracy but they all present challenges. One such technique is the adoption of the maximum-disc hypothesis (Van Albada et al. 1985). This method involves adopting a $M/L$ such that there is maximum contribution from the disc without exceeding the observed rotation curve. However, there is still argument about whether this hypothesis is correct. Another technique used to estimate the $M/L$ is from stellar population synthesis models. However, this method involves several significant assumptions about the star formation and chemical enrichment histories and the initial stellar mass function, and it needs an accurate account of late phases of stellar evolution (Maraston

⋆ Email: suryashree.aniyan@anu.edu.au

© 0000 The Authors

2005; Conroy et al. 2009). The *M/L* obtained using these methods (in the K-band) have typical uncertainties of ∼ 0.3 dex (see for e.g. Conroy 2013; Courteau et al. 2014), enough to allow a maximal or sub-maximal solution in most mass modeling decompositions.

One of the more direct methods to break the disc-halo degeneracy uses the vertical velocity dispersion of tracers in the discs to measure the surface mass density of the disc (e.g. Van der Kruit & Freeman 1984, Bottema et al. 1987, Herrmann et al. 2008, Bershady et al. 2010a). Using the 1D Jeans equation in the vertical direction, the vertical luminosity-weighted velocity dispersion $\sigma_z$ (integrated vertically through the disc) and the vertical exponential disc scale height $h_z$ together give the surface mass density $\Sigma$ of the disc via the relation:

$$\Sigma = f\sigma_z^2/Gh_z \quad (1)$$

where G is the gravitational constant and $f$ is a geometric factor, known as the vertical structure constant, that depends weakly on the adopted vertical structure of the disc. For example, for an isothermal disc with $\rho(z) \propto \mathrm{sech}^2(z/2h_z)$, the factor $f = f_{iso} = 1/2\pi$, whereas $f = f_{exp} = 2/3\pi$ for a vertically exponential disc with $\rho(z) \propto \exp(-z/h_z)$ (Van der Kruit & Freeman 2011). Van der Kruit (1988) advocated for an intermediate case where $\rho(z) \propto \mathrm{sech}(z/h_z)$, for which $f = f_{int} = 2/\pi^2$. Thus, having adopted a vertical structure for the stellar disc, we need two observables to estimate the surface mass density of the disc: the scale height and the vertical velocity dispersion. The surface brightness of the disc and the surface mass density ($\Sigma$ from equation 1) together give the *M/L* of the disc, which is needed to break the disc-halo degeneracy.

The scale height $h_z$ of the thin disc is typically about 300 pc (see for e.g. Gilmore & Reid 1983), but cannot be measured directly for face-on galaxies. Studies of edge-on disc galaxies show a correlation between the scale height and indicators of the galaxies' mass scale, such as the absolute magnitude and the circular velocity. Yoachim & Dalcanton (2006) show the correlation of the scale heights of the thin and thick disc with circular velocity of edge-on disc galaxies using R-band surface photometry. Similarly, Kregel et al. (2005) used I-band surface photometry of edge-on disc galaxies to derive correlations between the scale height and intrinsic properties of the galaxy such as its central surface brightness. We can, therefore, estimate the scale height statistically using other known features of the galaxy.

The other parameter, the vertical stellar velocity dispersion $\sigma_z$ of the disc, can be measured in relatively face-on galaxies from:

- spectra of the integrated light of the disc.
- the velocity distribution of a population of stellar tracers (such as planetary nebulae).

Using the integrated light to measure $\sigma_z$ is challenging because high resolution spectra of low surface brightness discs are required to measure the small velocity dispersions (e.g. for the old disc near the sun, Aniyan et al. (2016) find $\sigma_z \sim$ 20 km s$^{-1}$). Another challenge comes from the fact that near face-on galaxies are rare, so dynamical analyses are required in galaxies with larger inclinations to extract the vertical component $\sigma_z$ from the observed line-of-sight velocity dispersion (LOSVD) $\sigma_{LOS}$. NGC 628 is one of the few galaxies (the only one in our sample), which is so nearly face-on that the in-plane components of the stellar motion makes a negligible contribution to the LOSVD. Van der Kruit & Freeman (1984), Bottema et al. (1987) and Bershady et al. (2010a) have used this method and find that the disc *M/L* is relatively low and the discs are submaximal.

The DiskMass Survey (DMS; Bershady et al. 2010a) used integral-field spectroscopy to measure the stellar kinematics of the discs of near face-on galaxies observed with the SparsePak and PPak instruments. The DMS measured stellar kinematics for 46 galaxies and calculated their vertical velocity dispersions from the absorption line spectra of the integrated disc light. They then combined these dispersions with the estimated scale heights to calculate the surface mass density of the disc (using equation 1). Bershady et al. (2011) find that the dynamical stellar *M/L* obtained from the surface mass density is about 3 times lower than the *M/L* from the maximum disc hypothesis and conclude that discs are submaximal.

Herrmann et al. (2008) and Herrmann & Ciardullo (2009a,b) observed 5 near-face-on spirals (including our target galaxy NGC 628) using PNe as tracers. The advantage of using PNe as tracers over integrated light work is that it enables one to extend the analysis to the outer regions of the disc. Herrmann & Ciardullo (2009b) find that 4 of their discs appear to have a constant *M/L* out to ∼ 3 optical scale lengths. Beyond this radius, $\sigma_z$ flattens out and remains constant with radius. Herrmann & Ciardullo (2009b) suggest that this behaviour could be due to an increase in the disc mass-to-light ratio, an increase in the contribution of the thick disc, and/or heating of the thin disc by halo substructure. They also find a correlation between disc maximality and whether the galaxy is an early or late type spiral. They note that the later-type (Scd) systems appear to be clearly submaximal, with surface mass densities less than a quarter of that needed to reproduce the central rotation curves, whereas in earlier (Sc) galaxies (like NGC 628), this discrepancy is smaller, but still present; only the early-type Sab system M94 has evidence for a maximal disc (Herrmann & Ciardullo 2009b).

An important conceptual problem has, however, been overlooked in the earlier studies described above. Equation 1 comes from the vertical Jeans equation for an equilibrium disc. It is therefore essential that the vertical disc scale height $h_z$ and the vertical velocity dispersion $\sigma_z$ should refer to the same population of stars.

The red and near-infrared measurements of the scale heights of edge-on disc galaxies are dominated by the red giants of the older, kinematically hotter population. The dust layer near the Galactic plane further weights the determination of the scale height to the older kinematically hotter population: e.g. De Grijs et al. (1997). On the other hand, the velocity dispersion $\sigma_z$ is usually measured from integrated light spectra near the Mg b lines (∼ 5150 − 5200 Å), since this region has many absorption lines and the sky is relatively dark. The CaII triplet region at ∼ 8500 Å is also a potential region of interest with several strong absorption features. However, there are many bright sky emission lines in this region, which makes the analysis more difficult. The CaII triplet wavelength regions are also affected by Paschen lines from young hot stars and are not dominated by the red giants alone (see Figure 6 and associated discussion in Iodice et al. 2015). The discs of the gas-rich galaxies for which good HI rotation data are available usually have a continuing history of star formation and therefore include a population of young (ages < 2 Gyr), kinematically cold stars among a population of older, kinematically hotter stars. The red giants of this mixed young + old population provide most of the absorption line signal that is used for deriving velocity dispersions from the integrated light spectra of galactic discs.

Therefore, in equation 1, we should be using the velocity dispersion of the older disc stars in combination with the scale heights of this same population for an accurate determination of the surface mass density (Jeans 1915). In practice, because of limited signal-to-noise ratios for the integrated light spectra of the discs, integrated light measurements of the disc velocity dispersions usually adopt a single kinematical population for the velocity dispersion whereas, ideally, the dispersion of the older stars should be extracted from the composite observed spectrum of the younger and older stars.

Adopting a single kinematical population for a composite kinematical population gives a velocity dispersion that is smaller than the velocity dispersion of the old disc giants (for which the scale height was measured), and hence underestimates the surface density of the disc. A maximal disc will then appear submaximal. This problem potentially affects the usual dynamical tracers of the disc surface density in external galaxies, like red giants and planetary nebulae, which have progenitors covering a wide range of ages. It therefore affects most of the previous studies. It is consistent with the discovery by Herrmann & Ciardullo (2009b), mentioned above, that the later-type (Scd) systems appear to be clearly submaximal, because these later-type systems are potentially the most affected by the contribution of the younger planetary nebulae to the velocity dispersion (see however Courteau et al. 2014 and Courteau & Dutton 2015). A recent study of the K-giants in the V-band in the solar neighbourhood by Aniyan et al. (2016) showed that the young stars contribute significantly to the total light and that the velocity dispersion derived assuming a single population of tracers (red giants) leads to the disc surface mass density being underestimated by a factor ∼ 2.

Our goal in this paper is to use the kinematics and scale height of the older stars as consistent tracers to estimate the total surface density of the disc (older stars + younger stars + gas). The distribution of the older stars will be affected by the gravitational field of the thinner layer of younger stars and gas. Their dynamical contribution is often neglected in estimates of the disc surface density. If we assume that the layer of younger objects and gas is very thin, and take the velocity distribution of older stars as isothermal, then there is an exact solution for the density distribution of the older stars (see Appendix A). Their density distribution is a modified version of the familiar sech$^2(z/2h_z)$ relation for the simple isothermal, and Equation 1 becomes:

$$\Sigma_T = \Sigma_D + \Sigma_{C,*} + \Sigma_{C,gas} = \sigma_z^2/(2\pi G h_z) \quad (2)$$

where $\Sigma_T$ is the total surface density of the disc, $\Sigma_D$ is the surface density of the older stellar component which we are using as the dynamical tracer (its scale height is $h_z$ and its integrated vertical velocity dispersion is $\sigma_z$). $\Sigma_{C,*}$ and $\Sigma_{C,gas}$ are the surface densities of the cold thin layers of young stars and gas respectively. An independent measurement of $\Sigma_{C,gas}$ is available from 21 cm and mm radio observations. We will see later (Table 6) that the contributions of the cold layers to the total surface density can be significant.

In this paper, we present our observations of our most face-on galaxy NGC 628 (M74) to extract a two component velocity dispersion for the motion of the hot and cold disc component independently. We combine velocity dispersion data from two sources: (1) an absorption line study of the integrated disc light using spectra from the VIRUS-W IFU instrument on the 107-inch telescope at McDonald Observatory, and (2) the velocity distribution of planetary nebulae observed using the planetary nebula spectrograph (PN.S) on the William Herschel Telescope. Section 2 describes the observations and data reduction for VIRUS-W, and section 3 summarises the same for the PN.S. Section 4 discusses the photometric properties and derives scale height of NGC 628 and section 5 briefly summarises the adopted parameters that goes into our analysis in the calculation of the surface mass density of the disc. Section 6 discusses our analysis to derive the surface mass density of the cold gas in this galaxy and section 7 details the analysis involved in the extraction of a double Gaussian model from our data. Section 8 discusses the vertical dispersion profile of the hot and cold stellar components, and section 9 describes the calculation of the stellar surface mass density. Section 10 explains the rotation curve decomposition using the calculated surface mass densities. Section 11 lists our conclusions and scope for future work. In the Appendix, we discuss the dynamical effect of the cold disc component on the hot component.

## 2 VIRUS-W SPECTROGRAPH

The VIRUS-W is an optical-fibre-based Integral Field Unit (IFU) spectrograph built by the University Observatory of the Ludwig-Maximilians University, Munich and the Max-Planck Institute for Extraterrestrial Physics, and used on the 2.7m Harlan J. Smith Telescope at the McDonald Observatory in Texas. The IFU has 267 fibres, each 150 $\mu$ m-core optical fibers with a fill factor of 1/3. With a beam of f/3.65, the core diameter corresponds to 3.2″ on sky, and the instrument has a large field of view of 105″ × 55″ (Fabricius et al. 2012). We use the high resolution mode of the instrument which has a spectral resolving power of R ∼ 8700 or an average velocity resolution of about 14.7 km s$^{-1}$ (gaussian sigma of the PSF). The spectral coverage is 4802 – 5470 Å. The instrument is ideally suited for the study of the absorption features in the Mgb region (∼ 5175 Å). We summed the spectra over the IFU, excluding those affected by foreground stars, to produce summed spectra of high signal-to-noise ratio (SNR) at two mean radii. The high SNR allows us to measure velocity dispersions somewhat lower than the velocity resolution (sigma) of the instrument.

### 2.1 Observations

NGC 628 is a large nearby galaxy, much larger than the field of the IFU. It was observed in October 2014. We were able to observe several fields around the galaxy with a luminosity weighted radius of about 78″. This corresponds to about 1 scale length in the R-band (Fathi et al. 2007; Möllenhoff 2004). We positioned the IFU along the major and minor axis as well as at intermediate position angles. Our IFU positions on the galaxy are shown in Figure 1. The distribution of fields around the galaxy allows us to separate the contributions to the line of sight velocity dispersion from the vertical and in-plane components of the stellar motions in the disc. Since the fields cover a large radial extent on the galaxy, we decided to split the data into two radial bins, at luminosity-weighted radii of 62″ and 109″ respectively.

The position and exposure time at each position is given in

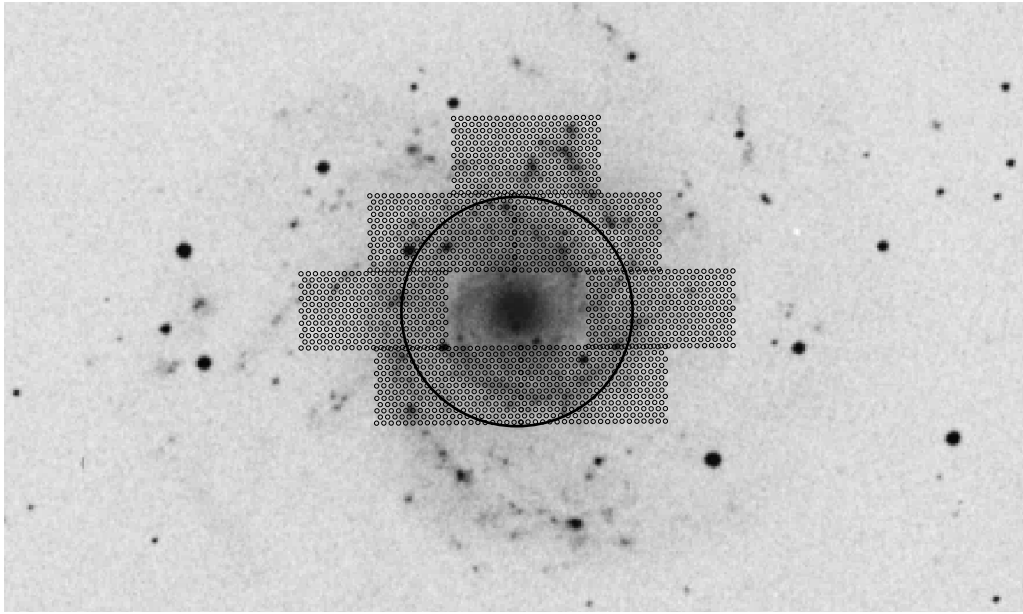

**Figure 1.** The positions of the VIRUS-W IFU fields overlaid on a DSS image of NGC 628. The position of the 267 fibres in each field are also shown. The circle at 85″ shows where we separated our data into the inner and outer radial bin.

Table 1. Each of the galaxy exposures were preceded and followed by a sky exposure of equal time. We repeated this sky -> galaxy -> sky sequence at least thrice at each field, as indicated in column 3 of Table 1. This enabled very good sky subtraction using the automated pipeline developed for VIRUS-W.

| RA (J2000) | Dec (J2000) | Exposure Time (s) |
|---|---|---|
| 1:36:49.00 | +15:47:02.7 | 3 × 800 |
| 1:36:34.36 | +15:47:02.1 | 3 × 800 |
| 1:36:45.19 | +15:47:02.1 | 3 × 800 |
| 1:36:37.85 | +15:46:06.5 | 3 × 800 |
| 1:36:45.46 | +15:48:00.0 | 3 × 800 |
| 1:36:38.10 | +15:47:59.1 | 3 × 800 |
| 1:36:41.16 | +15:48:56.9 | 5 × 800 |

**Table 1.** Coordinates and exposure times for the IFU fields in NGC 628.

### 2.2 Data Reduction and Extraction of Spectrum

The raw data were reduced using the automated pipeline 'CURE' which was orginally developed for HETDEX, but later adapted for VIRUS-W data reductions. The pipeline uses the biases and dome flats obtained during observation to debias and flat field correct the raw data. The pipeline then uses the observed arc frames for the wavelength calibration of the images. The final step is extraction of the spectrum from each fibre and then subtracting the sky. The sky frames preceding and succeeding the galaxy image are averaged and scaled to match the exposure time of the galaxy frame, which is then subtracted from the galaxy image. The data were reduced in log-wavelength space. The velocity step of the spectrum is $\sim 11$ km s$^{-1}$. As a check on the stability of the instrument we independently measured the dispersion of a few arc lines. Our measured values agree with the dispersions quoted in Fabricius et al. (2012) with $\sigma \sim 14$ km s$^{-1}$ near the Mgb region. As an added check, we combined all of our sky images to produce a 2D sky image with very high counts. We then measured the wavelengths of some known sky emission lines in the 1D spectrum from one of the fibres in this 2D image and compared them with the Osterbrock et al. (1996) wavelengths. This comparison is shown in Table 2. Since the positions of the emission lines in this spectrum match the known values, we cross-correlated the other 266 fibre spectra with this spectrum to see if there are any significant shifts in the wavelength solution. The shifts obtained from the correlation peak are all $< 2$ km s$^{-1}$. Thus the VIRUS-W is a very stable instrument and the errors in the wavelength system make a negligible contribution to the error budget.

The sky subtracted images from the reduction pipeline were combined and the spectra from each fibre in each field were summed to get a single spectrum at each of our two radial bins. The spectrum from each fibre was corrected for variations in systematic velocity over the IFU before they were summed together. This is explained in detail in section 7.1.

| Measured Wavelength (Å) | Osterbrock Wavelength (Å) |
|---|---|
| 5202.89 | 5202.98 |
| 5238.81 | 5238.75 |
| 5255.97 | 5256.08 |

**Table 2.** Comparison between the measured wavelengths of the sky lines from one of the fibres of our combined sky spectrum and the values from Osterbrock et al. (1996). This fibre was then cross-correlated with the other fibres to check for any significant wavelength shifts. The shifts were all $< 2$ km s$^{-1}$, indicating that errors in the wavelength system make a negligible contribution to the error budget.

### 3 PLANETARY NEBULA SPECTROGRAPH

Planetary nebulae (PNe) are part of the post-main-sequence evolution of most stars with masses in the range 0.8 to 8 M$_\odot$. Up to 15% of the flux from the central stars of PNe is reprocessed

*Resolving the Disc-Halo Degeneracy I* 5into the [OIII] emission line at 5007 Å (Dopita et al. 1992). These objects are plentiful in stellar populations with ages between 0.1 and 10 Gyr. The above properties make PNe useful probes of the internal kinematics of galaxies. They can be detected in galaxies out to many Mpc. They are easier to detect at large galactocentric radii where the background continuum is fainter, and are therefore an important complement to integrated light absorption-line studies.

The planetary nebula spectrograph (PN.S) is an imaging spectrograph designed for efficient observation of extragalactic PNe, and is used for the present project (Douglas et al. 2007). It operates on the 4.2 m William Herschel Telescope at La Palma, and has a field of view of 10.4 × 11.3 arcmin$^2$. The PN.S has a 'left' and 'right' arm in which the light is dispersed in opposite directions. Combining these two counter-dispersed images allows the PNe to be detected and their radial velocities to be measured in a single observation. The PN.S also has an undispersed H$\alpha$ imaging arm which can help to distinguish HII regions and background Ly$\alpha$ emitters from the PNe.

The PN.S is used by the PN.S collaboration, so far mainly on early-type galaxies (Coccato et al. 2009; Cortesi et al. 2013) plus a study of PNe in M31 (Merrett et al. 2006). Arnaboldi et al. (2017) describe a new survey of nearby face-on disc galaxies, aimed at measuring the internal kinematics of these discs, and illustrate the analysis of the new PN.S data for the prototypical galaxy NGC 628. The present paper presents the first results derived from these measurements in an attempt to break the disc-halo degeneracy.

### 3.1 Observations, Data Reduction, and Velocity Extraction

The data for NGC 628 were acquired over two nights during a 4 night observing run in September 2014. The weather during the run was excellent, with typical seeing being ∼ 1″. We obtained 14 images centred on the centre of the galaxy, each with an exposure time of 1800s. At the redshift of NGC 628, the wavelength of the [OIII] emission is near 5018 Å.

A detailed description of the data reduction can be found in Douglas et al. (2007) and Arnaboldi et al. (2017). The automated reduction procedure debiases and flat-field-corrects the raw images from the left and right arms, using bias frames and flats obtained during the observing run. Cosmic rays are removed using a custom-built routine in the pipeline. The wavelength calibration of the dispersed images was improved for this project by implementing a higher-order polynomial fit to the arc line calibration images taken during the observing run. After wavelength calibration, the 14 left and right arm images were stacked to create the final dispersed galaxy images.

Simultaneously with the [OIII] imaging, NGC 628 was also observed in H$\alpha$, using the H$\alpha$ narrow band filter on the undispersed H$\alpha$ arm of the PN.S. The H$\alpha$ arm and the reduction of these data are described in Arnaboldi et al. (2017).

### 3.2 Identification of Sources

Identification of PNe in late type galaxies brings in a new set of challenges, due mainly to contamination from HII regions. HII regions can also have strong [OIII] emission, and it is important to distinguish them from true PNe candidates.

Arnaboldi et al. (2017) describes the extraction of [OIII] emitters in the stacked left and right arm images for NGC 628.

After removing extended sources, we were left with a catalogue of 716 spatially unresolved [OIII] sources. From the measured positions of these sources on the left and right images, astrometric positions and LOS velocities were derived simultaneously.

We converted our instrumental magnitudes to the $m_{5007}$ magnitude scale used by Herrmann & Ciardullo (2009b), using our spectrophotometric calibration. This is accurate to within 0.05 mag. This allows us to directly compare our results to the values in Herrmann & Ciardullo (2009b). From here on, we shall only be using these $m_{5007}$ values. The bright luminosity cut-off for PNe in this galaxy is expected to be $m_{5007}$ = 24.73 (see Figure 2).

Our sample of 716 identified sources is still a mixture of spatially unresolved HII regions and PNe, since both can have strong [OIII] emissions. In the companion paper, Arnaboldi et al. (2017), we detail how we separated the spatially unresolved HII regions from PNe in the disc of NGC 628 using an [OIII]/H$\alpha$ color-magnitude cut that accounts for the apparent [OIII] magnitude of the bright cut-off in the PNLF and the large [OIII]/H$\alpha$ emission line ratio of bright PNe.

The line-of-sight velocity distributions of the HII regions and the PNe have different second moments ($\sigma_{LOS}$) in different radial bins. The $\sigma_{LOS}$ for the PNe correlates with m$_{5007}$. There is a kinematically cold population near the PNLF bright cut off, and then the velocity dispersion increases towards fainter magnitudes. This correlation is reminiscent of the age-magnitude-(vertical velocity dispersion) relation of the K-giant stars in the solar neighbourhood as shown by Aniyan et al. (2016) (also see Figure 10 in this paper).

Another possible source of contaminants in the emission line sample are historical supernovae. According to the IAU Central Bureau for Astronomical Telegrams (CBAT) List of Supernovae website (http://www.cbat.eps.harvard.edu/lists/Supernovae.html), there are three known historical supernovae in NGC 628. None of these objects made it into our PNe sample. An [OIII] emission line source was found at a distance of 1.6″ from SN 2002ap. However, on applying our colour-magnitude cut, this object was classified as an HII region. We, therefore, conclude that these contaminants are removed from our PNe sample by the colour-magnitude cut as well.

Figure 2 shows the luminosity function, including all 716 sources, and indicates the position of the bright luminosity cut-off for the PNe. The colour-magnitude cut on our 716 emission objects left us with about 400 objects. The LOS velocities for this sample are then used to calculate the velocity dispersions for the hot and cold PNe components.

### 3.3 Velocity errors

Space and wavelength information are closely related in an imaging spectrograph like the PN.S (see section 3.2). The left and right PN.S images are registered on the best quality image, which had the best seeing etc, so there is some correlation between the frames. In order to get an empirical estimate of the radial velocity measuring errors associated with each PN, we divided our 14 left and right images into two sets and then independently identified the unresolved [OIII] sources in each set. We had to split our sample into a set of 8 images and 7 images, since the 'reference image' used to stack the other images was common to both sets. We assume the radial velocity errors depend only on the total counts, not on the number of frames and

MNRAS **000**, 000–000 (0000)

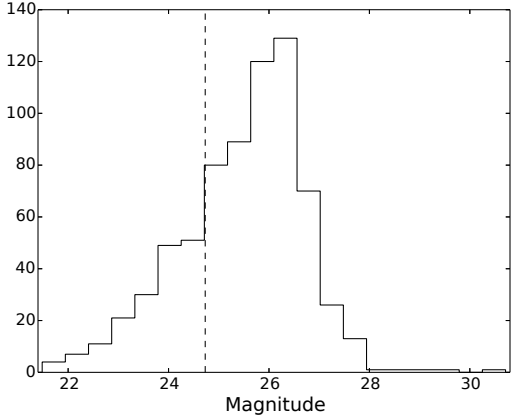

**Figure 2.** The luminosity function for all spatially unresolved [OIII] emitters identified in the combined left and right images of the PN.S. The dashed line shows the expected bright luminosity cut-off for PNe. We include only objects fainter than this value in our analysis. Objects brighter than the cut-off are mostly obvious bright HII regions.

that the velocity error of a single measurement at each count level, for all 14 frames, is the (rms of the difference between two velocity measurements at that count level)/$\sqrt{2}$, if there was no correlation between different images. However, since we had one image in common between the two sets, we carried out monte carlo simulations on the two image sets and the combined final image and found that the typical radial velocity error of a single measurement at each magnitude in the final image is (1/1.805) times the rms velocity difference between the two image sets at the same magnitude. Figure 3 shows the error expected for a single measurement from the whole set of 14 images, as a function of the $m_{5007}$ magnitude. Objects used in the subsequent analysis are those to the right of the vertical dashed line, which marks our bright cut-off. Most of these objects have estimated radial velocity errors between about 4 and 9 km s$^{-1}$.

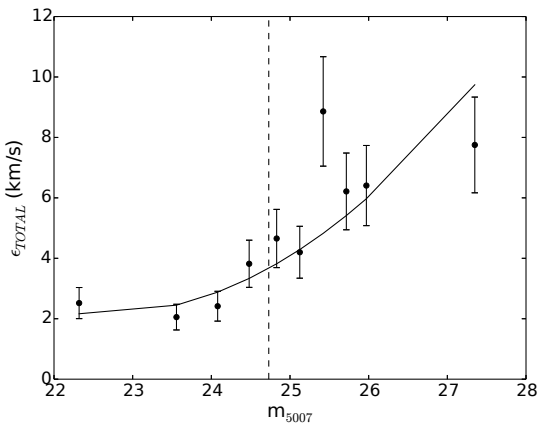

**Figure 3.** The measuring error for our sample of objects as a function of apparent magnitude. The magnitude system is the same as in Herrmann & Ciardullo (2009b). The dashed line shows the bright luminosity cut-off for this galaxy. Only the objects fainter than this magnitude were used in our analysis. The solid curve is the best fit to the data.

## 4 PHOTOMETRIC PROPERTIES AND SCALE HEIGHT

We use BVRI surface brightness profiles from Möllenhoff (2004), consistent with the Herrmann & Ciardullo (2009b) analysis. Möllenhoff (2004) tested their fit procedures extensively with artificial galaxies, including photon noise and seeing convolution. The statistical errors were found to be very small. The relevant errors were the systematic errors like the non-correct sky-subtraction, non-uniformness of the sky, errors in the determination of the seeing point-spread-function (Möllenhoff 2004). To estimate the error contributions of these effects, artificial galaxy images with typical sky levels, shot noise and seeing convolution were fitted with their 2-dimensional models. The sky level and the PSF were artificially set to different, slightly wrong values and the effect in the resulting photometric parameters was studied. They conclude that the errors due to inaccurate sky levels or PSF determinations are $\sim 5\%$ for the basic photometric parameters i.e the central flux density and the scale lengths (Möllenhoff 2004). We will adopt this error estimate in our analysis.

Determining the scale height for a face-on disc like NGC 628 is challenging. We need to make use of previous studies of edge-on discs that find correlations between the scale height and other properties of the galaxy such as its circular velocity (Yoachim & Dalcanton 2006) or its I-band scale length (Kregel et al. 2002). Since NGC 628 is so nearly face-on, it is difficult to measure its circular velocity $V_c$ directly. We attempted to make an independent estimate of the scale height, using the absolute magnitude of NGC 628 to estimate its circular velocity and hence the scale height. We used HI data for NGC 628 from the THINGS survey to determine $V_c = 180 \pm 9$ km s$^{-1}$. Our analysis for determining the rotation curve is detailed later in section 10. Yoachim & Dalcanton (2006) find the scale heights $h_z$ of the thin disc and circular velocities of edge-on galaxies (see Figure 9 in Yoachim & Dalcanton 2006) follow the relation $h_z = 305(V_c(\text{km s}^{-1})/100)^{0.9}$ pc. This study took the vertical density distribution to be isothermal. We use this relation to estimate $h_z = 518 \pm 23$ pc, which is much higher than the scale height of the MW $\sim 300$ pc. Yoachim & Dalcanton (2006) mention that for massive galaxies with large circular velocities ($V_c > 170$ km s$^{-1}$), their derived value for the scale height of the thin disc is larger than that for the MW. This could be because these galaxies have more prominent dust lanes, which may substantially obscure our view of the thin disc and lead to an overestimate of its scale height. Since the method described in Yoachim & Dalcanton (2006) is known to be uncertain for large dusty galaxies, we attempt to derive the scale height via alternate methods.

Herrmann & Ciardullo (2009b) reason that the scale height for NGC 628 should be in the range 300 − 500 pc based on the $h_z$ values obtained based on correlations of scale height with Hubble type (De Grijs & van der Kruit 1996), scale length (Kregel et al. 2002), and K-band central surface brightness of the galaxy (Bizyaev & Mitronova 2002). They further argue that for the thin stellar disc to be stable against axisymmetric perturbations, it should satisfy the Toomre (1964) criterion: $\sigma_R > 3.36G\Sigma/k$, where $\sigma_R$ is the radial component of the dispersion, $G$ is the gravitational constant, $\Sigma$ is the surface mass density of the disc, and $k$ is the epicycle frequency. Factoring these constraints into their analysis, they claim that $h_z = 400 \pm 80$ pc is a reasonable estimate of the scale height of NGC 628. However, disc stabil-ity arguments are not very well-established and have significant uncertainties associated with them.

Kregel et al. (2002) studied edge-on galaxies in the I-band and found correlations between the scale height and the I-band scale lengths. Using the redder I-band photometry minimises the effect of dust in the galaxy, while at the same time minimising the effects of PAHs that are a problem in the NIR wavelengths. Bershady et al. (2010b) fit the Kregel et al. (2002) data and find the relation: $\log(h_R/h_z) = 0.367 \log(h_R/\text{kpc}) + 0.708 \pm 0.095$. Using this relation for NGC 628, and adopting the I-band $h_R = 73.4 \pm 3.7''$ from Möllenhoff (2004) and distance = $8.6 \pm 0.3$ Mpc (Herrmann et al. 2008), we get $h_z = 397.6 \pm 88.3$ pc.

However, we could not access the surface brightness profile data from Möllenhoff (2004). We only had the central surface brightness and scale length of the fit to the data in the various bands. In order to verify that the scale lengths from Möllenhoff (2004) was reasonable, we decided to check the 3.6 $\mu m$ surface brightness profile for NGC 628 from the S4G survey (Muñoz-Mateos et al. 2013; Salo et al. 2015). Figure 4 shows the surface brightness profile of NGC 628 at 3.6 $\mu m$. It is clear from the figure that NGC 628 has a pure exponential disc with the scale length $h_{3.6} = 69.34''$ (Salo et al. 2015). The 3.6 $\mu m$ scale length agrees fairly well with the I-band scale length from Möllenhoff (2004). The red and green lines in Figure 4 are the fits to the bulge and disc respectively. While the total bulge light contributes only 6.5% to the total light of the galaxy, the bulge light dominates within the central 1.5 kpc and, therefore, it needs to be taken into account in the mass modelling.

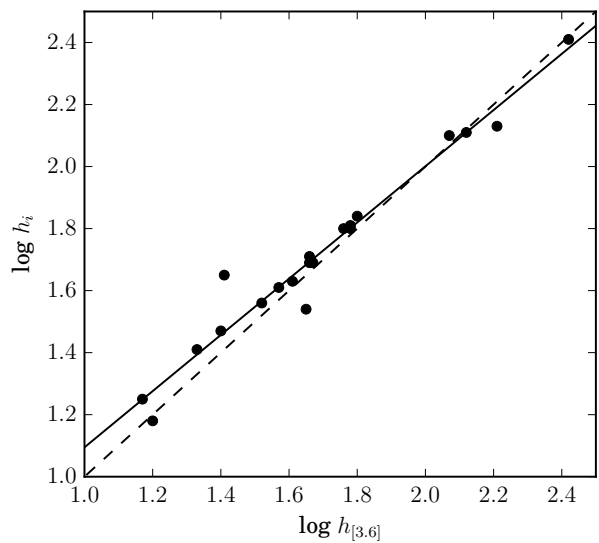

**Figure 5.** Relationship between the SDSS i-band scale length and the 3.6 $\mu m$ scale length from Ponomareva (2017). The solid line is a linear fit to the data. The dashed line is a line with slope = 1.

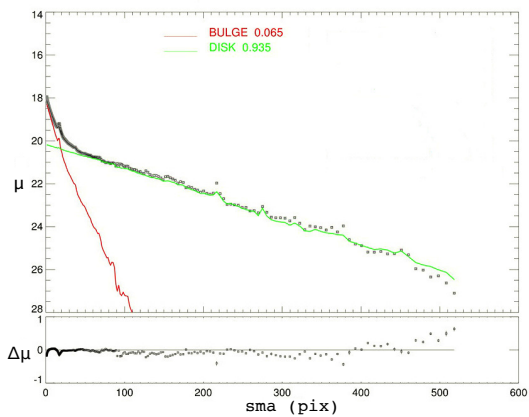

**Figure 4.** 3.6 $\mu m$ surface brightness profile from the S4G survey (Muñoz-Mateos et al. 2013; Salo et al. 2015). The y-axis shows the surface brightness profile (in AB magnitude) and the x-axis is the distance along the semi-major axis (with a pixel scale of 0.75 arcsec/pixel). The bottom panel shows the residuals between the data and the fit. The red and green lines are the fits to the bulge and exponential disc respectively. The bulge contributes only 6.5% of the total light in this galaxy.

The relation from Kregel et al. (2002) uses the I-band scale length. Having accurately determined the $h_{3.6}$ from Salo et al. (2015), we use the relation from Ponomareva (2017) between the scale lengths in i-band and 3.6 $\mu m$ band, calibrated for a sample of 20 disc galaxies. This relation is shown in Figure 5. This gives us the scale length in i-band via the relation: $\log(h_i) = 0.9 \log(h_{3.6}) + 0.19 \pm 0.05$. This gives us the scale length in the SDSS i-band as $70.3 \pm 8.1''$, which is close to the I-band scale length from Möllenhoff (2004). Using this value for the i-band scale length, gives us a scale height $h_z = 386.9 \pm 89.6$ pc.

The scale height obtained using the Möllenhoff (2004) photometry is remarkably close to the scale height estimate got using the 3.6 $\mu m$ photometry. We will therefore use the Möllenhoff (2004) photometry in all further analysis, and adopt the scale height value as $h_z = 397.6 \pm 88.3$ pc.

## 5 ADOPTED PARAMETERS

In order to proceed with the calculation of the surface mass densities and the subsequent $M/L$ of the disc, we need to establish the values that we will adopt for certain parameters. These parameters are obtained from previous literature values and are listed in Table 3.

The stellar velocity ellipsoid parameter, $\sigma_z/\sigma_R$, is rather uncertain for external galaxies. However, it is important to adopt a value for this parameter in order to convert our observed line-of-sight velocity dispersions to the vertical velocity dispersion. Solar neighbourhood studies have estimated this parameter to be between 0.5 – 0.7 (see Wielen 1977; Woolley et al. 1977; Bienaymé 1999; Dehnen & Binney 1998 ). Van der Kruit & de Grijs (1999) studied a sample of edge-on spiral galaxies and estimated their typical $\sigma_z/\sigma_R$. This analysis involves several dynamical assumptions and scaling arguments. They do not find any trend in $\sigma_z/\sigma_R$ as a function of morphological type or rotational velocity of the galaxy. Shapiro et al. (2003) studied six nearby spiral galaxies and combined their data with the results from Van der Kruit & de Grijs (1999). They find a marginal trend of a declining $\sigma_z/\sigma_R$ with Hubble type. However, these results have significant errors. For later type spirals, it can be argued that the $\sigma_z/\sigma_R$ doesn't show any trend, and seem to have a constant value of $\approx 0.6$ albeit with large uncertainties (see Figure 5 in Shapiro et al. 2003). We, therefore, adopt the $\sigma_z/\sigma_R$ to be $0.60 \pm 0.15$ (uncertainty at 25%) for this galaxy. This is similar to the value adopted by the DMS team (Bershady et al. 2010b). It is interesting to note that the error on this stellar velocity ellipsoid parameter has a negligible effect on the total er-ror budget for a galaxy as face-on as NGC 628 (see Figure 5 in Bershady et al. 2010b).

| Parameters | Value/ Description | Data source |
|---|---|---|
| Inclination | 8.5° ± 0.2° | Walter et al. (2008) |
| Distance | 8.6 ± 0.3 Mpc | Herrmann et al. (2008) |
| Scale length (I-band) | 73.4 ± 3.7″ | Möllenhoff (2004) |
| Scale height | 397.6 ± 88.3 pc | Kregel et al. (2002) |
| $\sigma_z/\sigma_R$ | 0.60 ± 0.15 | |
| Photometry | BVRI bands | Möllenhoff (2004) |
| Photometry | 3.6 $\mu m$ band | Salo et al. (2015) |

**Table 3.** The parameters for NGC 628 adopted from the literature and used in our analysis.

The inclination was determined via kinematic fit to the HI data from the THINGS survey (Walter et al. 2008). This procedure is detailed in section 10.

## 6 SURFACE MASS DENSITIES OF THE COLD GAS

As mentioned in section 1, our velocity dispersion analysis gives the total surface density of the disc, including the gas. We do however need the surface density of the cold gas to derive the separate surface densities of the hot and cold stellar components (see Table 6), because these components have different flattenings which should, for completeness, be included when computing their contributions to the rotation curve. We derived the HI surface density profile using the THINGS HI data for NGC 628 (Walter et al. 2008). We created an integrated column-density HI map by summing the primary beam-corrected channels of the clean data cube. The radial surface density profile was then derived by averaging the pixel values in the concentric ellipses projected on to the HI map. We will use the same radial sampling, position and inclination for obtaining the rotation curve (see section 10.1).

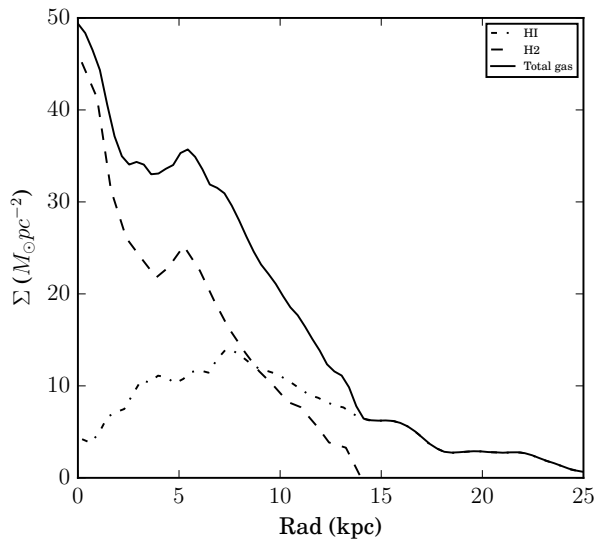

**Figure 6.** Surface mass density of the cold gas in NGC 628. The HI density profile from Walter et al. (2008) is shown as the dot dashed line and the H2 profile derived using the CO profile from Leroy et al. (2009) is shown as the long dashed curve. The surface density profile of the total gas is shown as the solid curve.

The resulting pixel values were converted from flux density units [Jy/beam] to column densities [atoms cm$^{-2}$], using equation 5 in Ponomareva et al. (2016). The resulting HI surface density profile is shown in Figure 6 in the dot dashed line. We adopted the error on the surface density as the difference between surface density profiles of the approaching and receding sides of the galaxy.

We derived the H2 surface density profile by using the CO profile from the HERACLES survey (Leroy et al. 2009). We then converted the CO intensities into H2 surface densities following the method outlined in Leroy et al. (2009). The resulting H2 profile is shown in Figure 6 as the long dashed curve. The error on the H2 densities were obtained from the HERACLES error maps for NGC 628.

The HI and H2 surface mass density profiles gives us the total gas surface mass density in this galaxy. This is shown as the solid line in Figure 6. All profiles were de-projected so as to be face-on and were corrected for the presence of helium and metals.

## 7 EXTRACTING VELOCITY DISPERSIONS OF THE HOT AND COLD COMPONENTS

### 7.1 Stellar Absorption Spectra

#### 7.1.1 Removing Galactic Rotation

For the VIRUS-W data, the automated pipeline 'CURE' returns a two dimensional FITS image, where each row represents a fibre spectrum and the x-axis is the wavelength dimension. Our goal is to measure the line of sight velocity dispersion ($\sigma_{LOS}$) without including the effects of galactic rotation across the field of the IFU. One option for removing galactic rotation would be to model the rotation field over the IFU using the observed rotation curve. Alternatively, we could use the local observed HI velocity at the position of each of the IFU fibres, and we have chosen this option. We used the 21 cm HI data from the THINGS survey (Walter et al. 2008). We assume that the spectrum from each fibre is shifted in velocity by the local HI velocity. Although this procedure removes the galactic rotation and any large scale streaming motions across the field of the IFU, it will however introduce an additional small component of velocity dispersion to the apparent stellar velocity dispersion. This is because the motion of the gas is not purely circular and we will need to correct for its (small) effect on the derived stellar dispersion.

Initially, we made a double Gaussian analysis to derive the velocity dispersion for the hot and cold components of the disc. To get sufficient SNR for this double Gaussian analysis, we sum up all the shifted spectra (one from each fibre) over the IFU field, to get a single spectrum (at each radial bin). IFU fibres that fell on stars in the field are excluded from the sum. We used the penalized pixel-fitting code pPXF developed by Cappellari & Emsellem (2004) (see also Coccato et al. 2011; Cappellari 2017) to get the mean velocity and velocity dispersion of the two components. This code uses a list of stellar templates to directly fit the spectrum in pixel space to recover the line of sight velocity distribution (LOSVD). pPXF can fit up to six higher moments to describe the LOSVD. It has options to fit either 1 or 2 LOSVD to the given spectrum, each with up to 6 moments. We used stars of different spectral types observed with VIRUS-W as our list of stellar templates. This avoids any problems of reso-lution mismatch between the stellar templates and the galaxy spectrum. pPXF then finds a best-fit spectrum to the galaxy spectrum, which is a linear combination of different stellar templates. We assume the two components of the LOSVD to be Gaussian for this nearly face-on galaxy, and therefore retrieved only the first and second moment parameters from pPXF.

The final summed spectra used in our analysis have an SNR of 79 and 62 per wavelength pixel for the spectrum from the inner and outer radial bins respectively (each wavelength pixel is ∼ 0.19 Å). These SNR values are empirical estimates obtained by taking into consideration the contribution of the galaxy and sky shot noise and the readout noise of the detector. The VIRUS-W instrument has a wavelength-dependent resolution, offering the highest resolution R ∼ 9000, around the Mgb region ($\lambda \sim 5160$ Å). Therefore, we only used the region between wavelengths of about 5050 – 5300 Å in our analysis, since it has the highest resolution and avoids the emission lines at lower wavelengths. The [NI] doublet emission lines from the interstellar medium of the galaxy can be seen at ∼ 5200 Å (see Figure 7). These are not residual sky lines: they appear at the redshift of the galaxy.

### 7.1.2 Measuring the LOSVD

As explained in section 7.1.1, we did a double Gaussian fit to the data, fitting for two moments for each component. In this case, pPXF returns the velocity and the line-of-sight (LOS) dispersions for the two component and for the single component fit. Adding more parameters to the model invariably improves the fit to the data. We therefore need to quantitatively decide whether the 2 Gaussian or single Gaussian model is a more appropriate fit to the data. To do this, we used the Bayesian Information Criterion (BIC; Schwarz 1978), which is calculated using the relation:

$$\text{BIC} = -2 \cdot \ln \hat{L} + k \cdot \ln(n), \quad (3)$$

where $\hat{L}$ is the maximized value of the likelihood function of the model, $n$ is the number of data points or equivalently the sample size and $k$ is the number of free parameters to be estimated. Under the assumption that the model errors are independent and are Gaussian, equation 3 becomes:

$$\text{BIC} = n \cdot \ln(RSS/n) + k \cdot \ln(n) \quad (4)$$

where RSS is the residual sum of squares.

The BIC penalises the model with the larger number of fitted parameters and, between 2 models, the model with the lower BIC value is preferred. The values of the BIC for our VIRUS-W spectra are tabulated in Table 4. Since the model with the lower value of BIC is preferred, the two component fit is preferred over the single component fits in both radial bins.

We then attempted to carry out a triple Gaussian fit to the data, to check if we have any contribution from the thick disc. However, we could not get a third component in our fit when the data were divided into two radial bins. The degeneracy between the hot thin disc and the thick disc component, led to errors that were unacceptably large. We were able to get a third component with a dispersion consistent with a thick disc component if we only considered one radial bin and summed up the data from all the fibres. However, this third component may just be an artefact of the gradient of the velocity dispersion, since we are summing up the data over such a large radial extent. The information criterion that we used to judge the best model also rejects the three component fit. Therefore, we conclude that there is no significant thick disc contribution present in our data.

pPXF found an excellent fit to our spectrum for the two-component case, as shown in Figure 7. It returns the adopted spectra of the individual components, and the 2 spectra that it returns are consistent with the spectra of red giants. The mean contributions of the cold and hot disc components to the total light are 36% and 64% respectively. Figure 8 compares the two components found by pPXF in the inner radial bin. These are a linear combination of unbroadened stellar spectra, identified by pPXF as the best fit to our galaxy spectrum. The colder component with the smaller dispersion (in red in Figure 8) are also weaker lined as compared to the hotter component. This shows that the colder component is in fact the younger of the two components.

As mentioned earlier, since we used the THINGS HI data to remove rotation across the fields, we need to correct these two dispersion values for the contribution from the scatter of the HI velocities about the mean smooth HI flow over the field of the IFU. This correction was determined by fitting a plane function $V = ax + by + c$ to the HI velocities at the $(x, y)$ location of the individual VIRUS-W fibres at each IFU pointing. The rms scatter of the HI velocities about this plane is 2.5 km s$^{-1}$. We note that this is the rms scatter of the mean HI velocities from fibre to fibre, which is not the same as the HI velocity dispersion. Correcting for this scatter changes the observed dispersions by only a very small amount. Table 4 shows our results after subtracting this value quadratically from the pPXF results.

The errors on the $\sigma_{LOS}$ are computed from monte carlo simulations. This was done by running 1000 iterations where, in each iteration, random Gaussian noise appropriate to the observed SN of the IFU data was added to the best fit spectrum originally returned by pPXF. pPXF was run again on the new spectrum produced in each iteration. The errors are the standard deviations of the distribution of values obtained over 1000 iterations. The errors on $\sigma_z$ presented in Table 4 take into account the errors on the inclination, $\sigma_z/\sigma_R$ value, as well as the monte carlo errors on $\sigma_{LOS}$. The errors on the LOS dispersions are the dominant source of errors.

| Mean Radius (arcsec) | 2 component Model | | | | 1 Component Model | | |
|---|---|---|---|---|---|---|---|
| | $\sigma_{z,cold}$ (km s$^{-1}$) | $\sigma_{z,hot}$ (km s$^{-1}$) | $\chi^2_{red}$ | BIC | $\sigma_z$ (km s$^{-1}$) | $\chi^2_{red}$ | BIC |
| 62 | 16.7 ± 3.6 | 55.4 ± 6.4 | 0.95 | 17923 | 31.9 ± 1.1 | 1.04 | 18107 |
| 109 | 15.2 ± 3.8 | 50.9 ± 8.9 | 1.11 | 19021 | 25.1 ± 1.2 | 1.15 | 19064 |

**Table 4.** The single and double Gaussian fit from pPXF. For each component, the Table gives the vertical velocity dispersion $\sigma_z$ for each of the components; see section 7.1.3. Dispersions have been corrected for the contribution from the HI velocity dispersion. An estimate of the reduced $\chi^2$ and the Bayesian Information Criterion parameter BIC defined in equation (3) is also given.

### 7.1.3 Extracting the Vertical Velocity Dispersion

The vertical component of the stellar velocity dispersion $\sigma_z$ was calculated from the line of sight component $\sigma_{LOS}$ by first calculating the azimuthal angle ($\theta$) to each fibre. The angle $\theta$ is measured in the plane of the galaxy, from the line of nodes. Then the LOS dispersion is given by:

$$\sigma^2_{LOS} = \sigma^2_\theta \cos^2\theta \cdot \sin^2 i + \sigma^2_R \sin^2\theta \cdot \sin^2 i + \sigma^2_z \cos^2 i + \sigma^2_{meas} \quad (5)$$

10 *S. Aniyan et al.*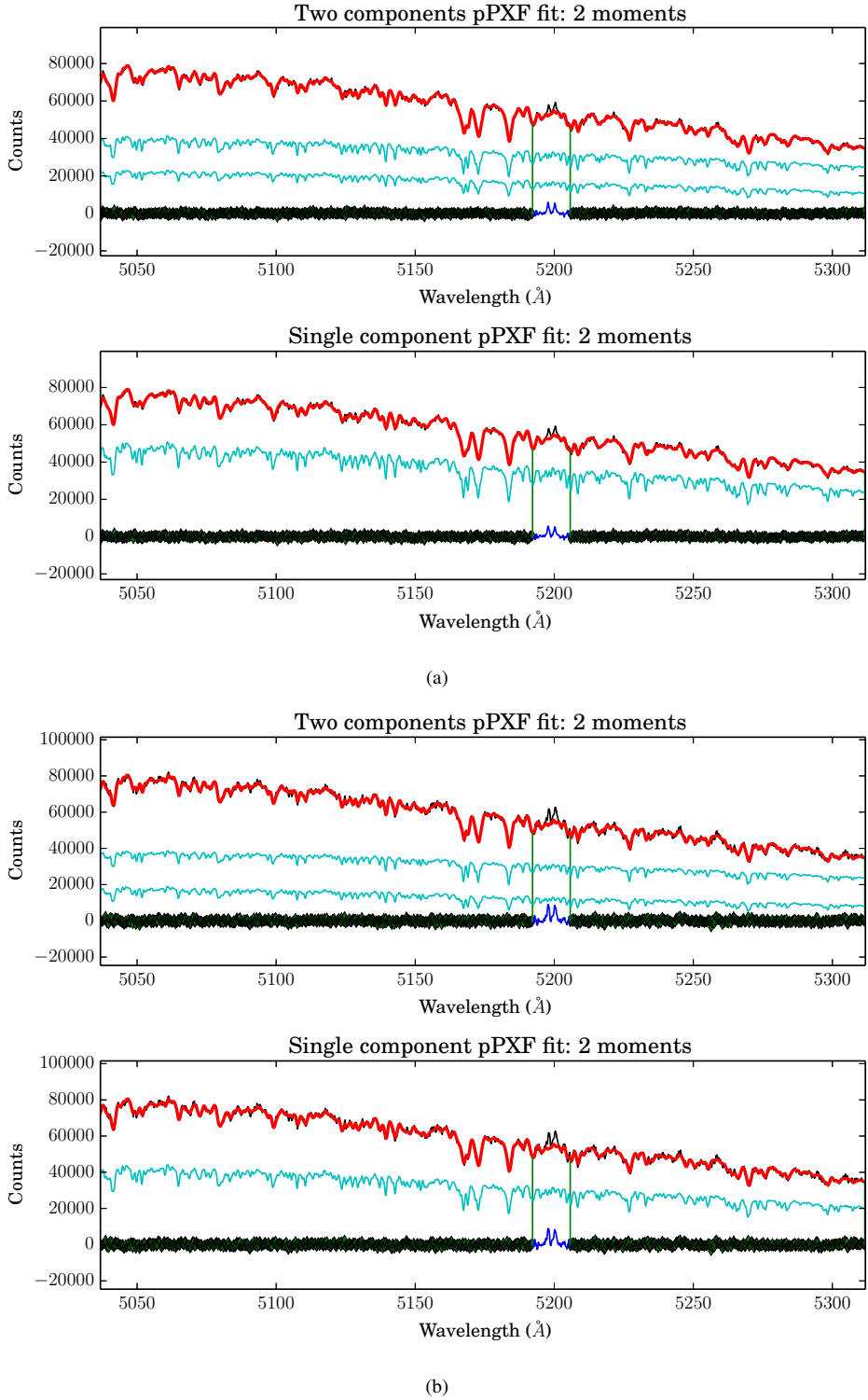

**Figure 7.** The pPXF fit results in (a) the inner radial bin at a luminosity weighted distance of 62″ and (b) the outer bin at a luminosity weighted distance of 109″. The upper panel shows the 2 component fit to the data whereas the lower panel shows a single component fit. Only the high resolution Mgb region of the spectrum was used for the fit. The galaxy spectrum is in black and the best fit from pPXF is in red. The cyan spectra are the two and one component spectra that pPXF found. The cyan spectra have been shifted vertically so as to be clearly visible. The residuals are shown in dark green. The [NI] doublet emission lines from the galaxy at ∼ 5200 Å have been omitted from the fit.

MNRAS **000**, 000–000 (0000)

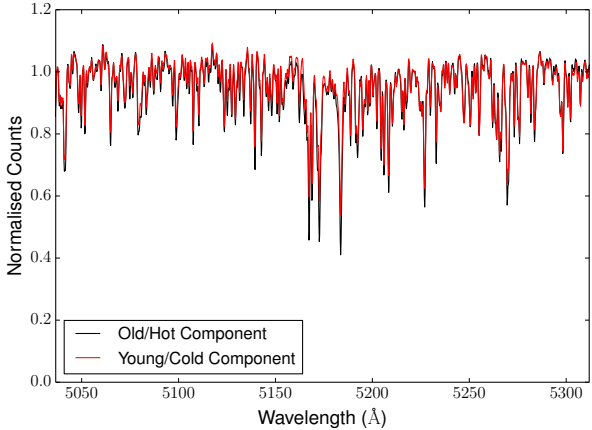

**Figure 8.** The two components found by pPXF in the inner radial bin of NGC 628. The spectrum in red represents the cold component, which is weaker lined than the hot component in black. The colder component found by pPXF is thus younger than the hotter component.

where $\sigma_R$, $\sigma_\theta$ and $\sigma_z$ are the three components of the dispersion in the radial, azimuthal and vertical direction, $\sigma_{meas}$ are the measurement errors on the velocity and $i$ is the inclination of the galaxy ($i = 0$ is face-on). This galaxy is too face-on to solve independently for the in-plane velocity dispersion components. We wish to remove the small contribution that the planar components make to the LOS distribution, so we adopt $\sigma_R = \sigma_\theta$ for $R < 80''$ where we take the rotation curve to be close to solid body. This is a fair assumption, based on an examination of the THINGS HI velocities along the galaxy's kinematic major axis. We also adopt the $\sigma_z/\sigma_R$ ratio to be $0.60 \pm 0.15$ (see Table 3), which is consistent with the value used by Bershady et al. (2010b) and the value found in the solar neighbourhood. Equation 5 then gives $\sigma_z$ in terms of $\sigma_{LOS}$. Since NGC 628 is almost face-on, the $\sigma_{LOS}$ and $\sigma_z$ values are almost the same.

Our choice to remove the rotation and streaming across the IFU fields by using the local HI velocities introduced a small additional broadening of the LOS velocity distribution, as described above. This small broadening ($\sim 2.5$ km s$^{-1}$) was quadratically subtracted from the dispersion values returned by pPXF. Our results for the stellar $\sigma_z$ values are presented in Table 4. The errors are the 1$\sigma$ errors from monte carlo simulations (as explained in section 7.1.2).

### 7.2 Planetary Nebulae

#### 7.2.1 Removing Galactic Rotation

As in the analysis of the IFU integrated light absorption spectra, we again need to remove the effects of galactic rotation from the PNe velocity field. We used the THINGS HI data as before, and obtained the HI velocity at the position of all our PNe from the THINGS first moment data. There appeared to be a small systematic offset $\sim 15$ km s$^{-1}$ between our PNe velocities and the THINGS HI data. We calculated this offset by cross-correlating the two data sets and determining the velocity of the correlation peak. This offset was then subtracted from the PNe velocities. The local HI velocities were then subtracted from these offset-corrected PNe velocities.

These velocities, corrected for the offset and with the galactic rotation removed are henceforth denoted $v_{LOS}$. They are the velocities that are used in our analysis to calculate the velocity dispersions. As for the IFU data (section 7.1.3), the radius and azimuthal angle ($\theta$) of the PNe in the plane of the galaxy were calculated, and the $v_{LOS}$ data were then radially binned into 3 bins, each with about 130 PNe. As explained in section 3.3, we applied a colour-magnitude cut using the [OIII] and H$\alpha$ magnitudes, to separate out the contamination from likely HII regions. Figure 9 shows the $v_{LOS}$ vs $\theta$ plots in each radial bin before and after the HI velocities were subtracted off.

#### 7.2.2 Extracting the LOSVD

In each radial bin, we remove a few 3$\sigma$ outliers, consistent with the analysis by Herrmann & Ciardullo (2009b), who clipped their sample of PNe to remove high-velocity contaminants from their sample. These outliers could be halo PNe or thick disc objects, and should be removed from our sample. Only a small number of objects in each radial bin have velocities $> 3\sigma$. A maximum likelihood estimator (MLE) routine written in python was then used to calculate the LOS velocity dispersions and the subsequent $\sigma_z$ in each radial bin.

The first iteration in this routine estimates $\sigma_{LOS}$ for the two components. The routine maximises the likelihood for the two-component probability distribution function given by:

$$P(\mu_1, \sigma_1, \mu_2, \sigma_2) =$$
$$\frac{1}{\sqrt{2\pi}} \left[ \frac{N}{\sigma_1} \exp\left(-\frac{(v_{LOS} - \mu_1)^2}{2\sigma_1^2}\right) + \frac{1-N}{\sigma_2} \exp\left(-\frac{(v_{LOS} - \mu_2)^2}{2\sigma_2^2}\right) \right]$$
(6)

In equation 6, $\mu_1$ and $\mu_2$ are the mean LOS velocities and $\sigma_1$ and $\sigma_2$ are the LOS dispersions of the cold and hot component respectively. $N$ is the fraction of the cold tracers in the data.

#### 7.2.3 Extracting the Vertical Velocity Dispersion

In order to calculate the surface mass density using equation 1, we need the vertical velocity dispersion of the hot component and the scale height of the same component. For NGC 628, which is a near face-on system, the $\sigma_z$ value will be very close to the $\sigma_{LOS}$ values. To determine this value, we again use an MLE method. Two parameters are passed to the function in this stage: $\sigma_{z1}$ and $\sigma_{z2}$ which are the vertical velocity dispersions of the cold and hot components respectively. The $\sigma_{LOS}$ values obtained using the method described above are passed to the routine as initial guesses, since the $\sigma_z$ will be very close to the value of $\sigma_{LOS}$ for this galaxy. We assume $f = \sigma_z/\sigma_R = 0.60 \pm 0.15$ and use inclination $i = 8.5° \pm 0.2°$ (see Table 3). The PN.S data are all at radii $> 80''$ where the rotation curve is flat, and we use the epicyclic approximation: $\sigma_R = \sqrt{2}\sigma_\theta$, where $\sigma_R$ and $\sigma_\theta$ are the in-plane dispersions in the radial and azimuthal directions. Now there is only one unknown $\sigma_z$, which we need to calculate.

Once the initial guesses are passed to the routine, it calculates the expected $\sigma_{LOS}$ for the hot and cold component at each azimuthal angle ($\theta$) using the relation:

$$\sigma_{LOS1}^2 = \frac{\sigma_{z1}^2 f^2}{2} \cos^2\theta . \sin^2 i + \sigma_{z1}^2 f^2 \sin^2\theta . \sin^2 i + \sigma_{z1}^2 \cos^2 i$$
$$\sigma_{LOS2}^2 = \frac{\sigma_{z2}^2 f^2}{2} \cos^2\theta . \sin^2 i + \sigma_{z2}^2 f^2 \sin^2\theta . \sin^2 i + \sigma_{z2}^2 \cos^2 i$$

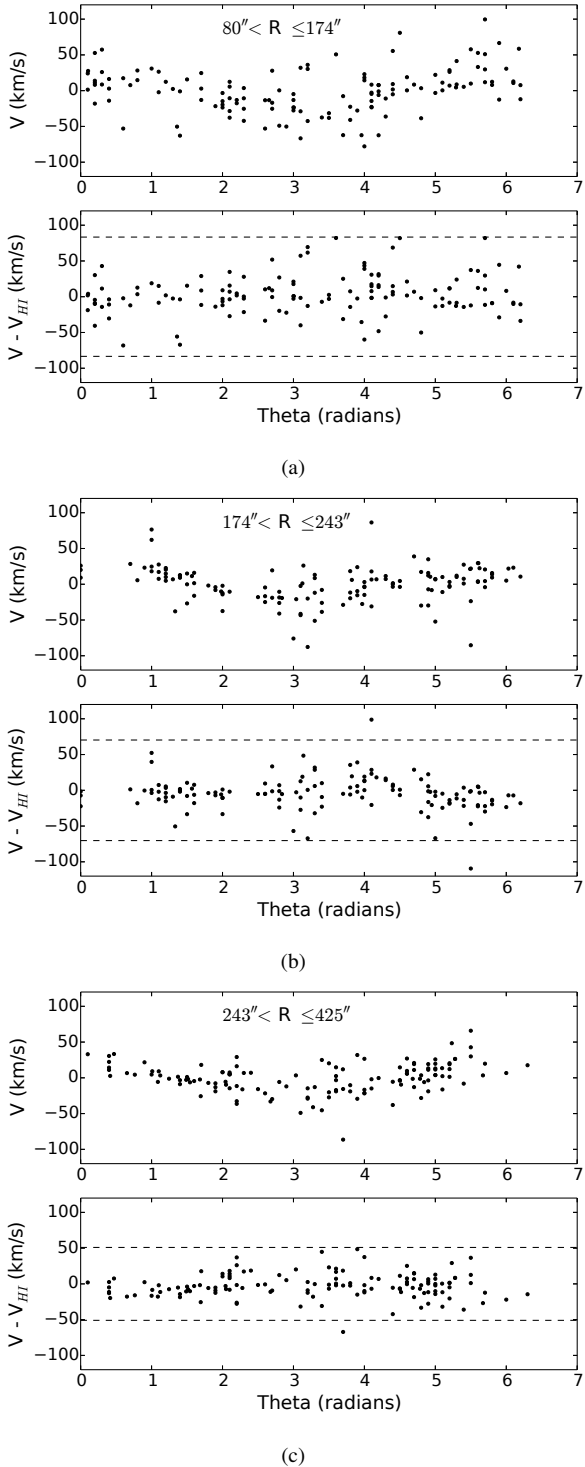

**Figure 9.** Velocity vs azimuthal angle plots in 3 radial bins, each with about 130 PNe. The angle $\theta$ is in the plane of the galaxy and measured from the line of nodes. The top panels show the velocity before correcting for galactic rotation and the bottom panels show the velocities after the THINGS HI velocities have been subtracted off. A few objects with velocity > $3\sigma$ were removed from the sample. The objects within the dashed lines are the ones that were included in our sample for analysis. Each panel shows a cold population of spatially unresolved emission line objects, plus a hotter population whose velocity dispersion appears to decrease with galactic radius.

The subscript 1 and 2 refers to the components of the cold and hot populations respectively. The code then proceeds to calculate the probability of a particular $v_{LOS}$ to be present at that azimuthal angle via the analytic equation:

$$P = \frac{N}{\sigma_{LOS1}\sqrt{2\pi}} \exp\left(\frac{-(v_{LOS} - \mu_1 \cos\theta . \sin i)^2}{2\sigma_{LOS1}^2}\right) + \frac{1-N}{\sigma_{LOS2}\sqrt{2\pi}} \exp\left(\frac{-(v_{LOS} - \mu_2 \cos\theta . \sin i)^2}{2\sigma_{LOS2}^2}\right) \quad (7)$$

In equation 7, $N$ is the value of the fraction of the cold population returned by the routine that calculated $\sigma_{LOS}$ described earlier. $\mu_1$ and $\mu_2$ were fixed at 0 km s$^{-1}$ in our analysis. We tried to leave the two means as parameters to be estimated by the code, but since NGC 628 has such a low inclination, these parameters would not converge. It is difficult to get an estimate of the asymmetric drift in such a face-on galaxy. So we assumed that the mean of the cold PNe were at the same velocity as the gas, at 0 km s$^{-1}$. We did try changing the mean of the hot PNe up to an asymmetric drift of 5 km s$^{-1}$ but this did not cause any significant changes in the $\sigma_z$ values. Equation 7 is then maximised to return the best fit values for $\sigma_z$ for the hot and cold population of PNe. The 1$\sigma$ errors $\sigma_z$ are calculated similar to the method used in the analysis of the VIRUS-W data. We carried out our monte carlo error estimation by using the double Gaussian distribution found by our MLE code, to pull out about 130 random velocities (i.e same number of objects as in each of our bins). The errors on the inclination and $\sigma_z/\sigma_R$ were also incorporated as Gaussian distributions in the simulation. We then used this new sample to calculate the $\sigma_z$ of the hot and cold component using our MLE routines. This whole process was repeated 1000 times, recording the dispersions returned in each iteration. The 1$\sigma$ error is then the standard deviation of the distribution of the dispersions returned from these 1000 iterations.

The parameters returned from the MLE routine that does the double Gaussian decomposition is given in Table 5. Figure 10 shows the $\sigma_z$ vs radius for NGC 628. The data points at a radius of 62″ and 109″ come from the VIRUS-W spectra in the inner field. Similar to the VIRUS-W analysis, we need to correct the dispersions for the PNe, because of the scatter in the the local THINGS HI velocity which we used to remove the galactic rotation. This correction was done as for the IFU data analysis by quadratically subtracting this small dispersion (∼ 2.5 km s$^{-1}$) from the values of the dispersions returned by the MLE routine. We also need to correct the measured dispersions for the measuring errors of the individual PNe, as shown in Figure 3. The rms measuring error in each radial bin is about 6 km s$^{-1}$. The formal variance of the corrected cold dispersion value at a radius of 208″ is negative. Hence we show its 90% confidence upper limit in Figure 10 and Table 5.

The PNe population selected by the colour-magnitude cut (see section 3.2) includes a population of kinematically cold sources. If this cold component had not been present, Herrmann & Ciardullo (2009b) would probably have measured larger velocity dispersions and higher surface mass densities. From the analysis of the cold and hot PNe population, as well as the identified HII regions (Arnaboldi et al. 2017), the PNe near the bright cut-off of the PNLF are kinematically colder than the HII regions at the same $m_{5007}$ magnitude, with $\sigma_{HII}/\sigma_{PNe}$ ∼ 2. Thus, these cold bright PNe are unlikely to be contaminant HII regions, since the LOS velocity distributions of the two classes of objects are so different. Work done by Miller Berto-lami (2016) suggests that PNe from massive progenitors evolve very fast. Thus massive progenitors, i.e. young massive stars, produce bright PNe that are still kinematically very cold.

## 8 VERTICAL VELOCITY DISPERSION PROFILE

Figure 10 shows the results obtained from the integrated light VIRUS-W data (points at R = 62″ and 109″) and the planetary nebulae from the PN.S data (3 outer points) in each radial bin. At each radius, we show our one component dispersion, and then the hot and cold thin disc dispersion from our double Gaussian fit. The dispersions obtained from Herrmann & Ciardullo (2009b) are also plotted for comparison.

The black square markers in Figure 10 show the radial dependence of the dispersion for the hotter old component of the stars and PNe, measured as described in the previous section. If the total surface density of the disc follows an exponential decline with radius, and the scale height of the old disc is constant with radius (as we are assuming), then we expect the vertical velocity dispersion to fall with radius, following $\sigma_z(R) = \sigma_z(0)\exp(-R/2h_{\rm dyn})$, where $h_{\rm dyn}$ is the scale length for the total (old + young stars + gas) surface mass density of the disc. We note that the photometric scale length varies with the photometric band, and $h_{\rm dyn}$ need not be equal to any of the photometric scale lengths. A fit of the above equation for $\sigma_z(R)$ gives the mass scale length $h_{dyn}$, which is also the scale length that should be used for calculating the rotation curve of the exponential disc (see section 10). Our fit of $\sigma_z(R)$ is shown as the solid curve in Figure 10: we find that the central velocity dispersion of the old disc population is $\sigma_z(0) = 73.6 \pm 9.8$ km s$^{-1}$ and the mass scale length $h_{dyn} = 92.7″ \pm 13.1″$, with significant covariance. The mass scale length is somewhat longer than the I-band scale length (Table 3), presumably because of the substantial contribution of the gas to the surface density at larger radii (see Table 6).

Figure 10 shows in red the PNe velocity dispersions derived for this galaxy by Herrmann & Ciardullo (2009b). Our one-component PNe velocity dispersions agree well with their results. We note that the difference between the one component dispersions and our hot component dispersions decreases with radius and is quite small for our outermost radial bin. This is consistent with the BIC values shown for the outer radial bin, which does not favour the two component model in the outer bin. The one component value for the last bin is also closer to our exponential disc curve than the two component value.

| Mean Radius | 2 component Model | | | 1 Component Model | |
|---|---|---|---|---|---|
| (arcsec) | $\sigma_{z,cold}$ (km s$^{-1}$) | $\sigma_{z,hot}$ (km s$^{-1}$) | BIC | $\sigma_z$ (km s$^{-1}$) | BIC |
| 132 | 4.6 ± 1.6 | 33.8 ± 3.3 | 1269 | 26.5 ± 1.8 | 1272 |
| 208 | ⩽ 6.7 ± 0.4 | 22.6 ± 2.1 | 1160 | 18.6 ± 1.3 | 1164 |
| 293 | 6.2 ± 1.7 | 17.5 ± 2.6 | 1124 | 14.5 ± 1.0 | 1118 |

**Table 5.** The $\sigma_z$ values calculated from the PN.S data. We give the 90% confidence upper limit for the cold dispersion in the second radial bin (see section 4.2.3 for details). The lower BIC values of the two component fit (except in the outermost radial bin) make it the preferred model over the one component model.

Having calculated the $\sigma_z$ in each radial bin in NGC 628, we can now proceed to calculate the surface mass density Σ of the disc using equation 2.

We now compare our results for the surface density of the disc, using a two-component (hot and cold) disc model, with the single-component analysis of Herrmann & Ciardullo (2009b). Herrmann & Ciardullo (2009b) adopted an intermediate vertical density distribution for their disc, with the geometric factor $f$ in equation 1 to be equal to $1/1.705\pi$. We adopt an isothermal model as described in Appendix A. A simple isothermal model with no additional cold layer has the $sech^2(z/2h_z)$ vertical density distribution. This distribution is exponential at large $z$ and flat near $z = 0$. In the presence of a significant cold layer, the $sech^2$ distribution is offset from zero, as shown in equation A7. For very small values, the offset parameter $b \sim \Sigma_C/\Sigma_T$ (this is true for our two inner radii). With a realistic surface density of kinematically cold gas and stars, the vertical distribution in equation A7 becomes close to exponential throughout. This effect is demonstrated in Figure 11 where we adopt an offset parameter of 0.5 and compare the sech$^2$ distribution with and without taking into account this offset. The offset distribution is close to an exponential beyond a height of ∼ 200 pc. This compares well with the study by Wainscoat et al. (1989) who found that the vertical surface brightness distribution of edge-on spirals in the NIR is close to exponential.

Using our scale height value $h_z = 397.6 \pm 88.3$ pc (see Table 3), which is almost the same as the value adopted by Herrmann & Ciardullo (2009b), and our $\sigma_z(0) = 73.6 \pm 9.8$ km s$^{-1}$ for the hot component, our total central surface mass density is $\Sigma(0) = 505 \pm 175$ M$_\odot$ pc$^{-2}$. This is a factor of 1.7× larger than the value calculated by Herrmann & Ciardullo (2009b). Note that we use an isothermal model of the vertical distribution as opposed to the intermediate model dopted by Herrmann & Ciardullo (2009b). Using the central surface brightness and scale lengths from Möllenhoff (2004) and Salo et al. (2015), we calculate the central luminosity of NGC 628 in units of L$_\odot$pc$^{-2}$. This is corrected for foreground extinction using Schlafly & Finkbeiner (2011) dust maps. Our central surface mass density divided by the calculated central luminosity gives the central $M/L$ in various photometric bands. The $M/L$ in the R band is found to be 2.0 ± 0.7, about 1.4× larger than the value of 1.4 ± 0.3 obtained by Herrmann & Ciardullo (2009b). Photometric errors of 5% have been included in the error estimate (Möllenhoff 2004; Muñoz-Mateos et al. 2013).

## 9 STELLAR SURFACE MASS DENSITY

### 9.1 Isothermal model including cold component

The equilibrium of the hot disc is determined by its own gravitational field plus the gravitational field of the cold layer (gas and young stars) which we can consider as external fields. The presence of these external fields changes the vertical structure of the old disc and affects the derived surface density and $M/L$ values for the old disc. We discuss this in more detail in Appendix A. We therefore use equation 2: $\sigma_z^2 = 2\pi G h_z \Sigma_T = 2\pi G h_z (\Sigma_C + \Sigma_D)$ to determine the surface mass density of the disc. We use the old disc population as a tracer of $\Sigma_T$, which is the parameter that determines the baryonic rotation curve.

The thin cold layer comprises gas and young stars, which we write as $\Sigma_C = \Sigma_{C,\,\rm gas} + \Sigma_{C,\,*}$ where $\Sigma_{C,\,\rm gas}$ is known directly from radio observations of the THINGS survey (Walter et al. 2008) and the HERACLES survey (Leroy et al. 2009) as discussed in section 6. The stellar contribution $\Sigma_{C,\,*}$ is not known directly from our data. In the following section, we estimate the effect of the cold gas layer at the five radii in NGC 628 (2.6 kpc to 12.2 kpc) where we have velocity dispersion data.

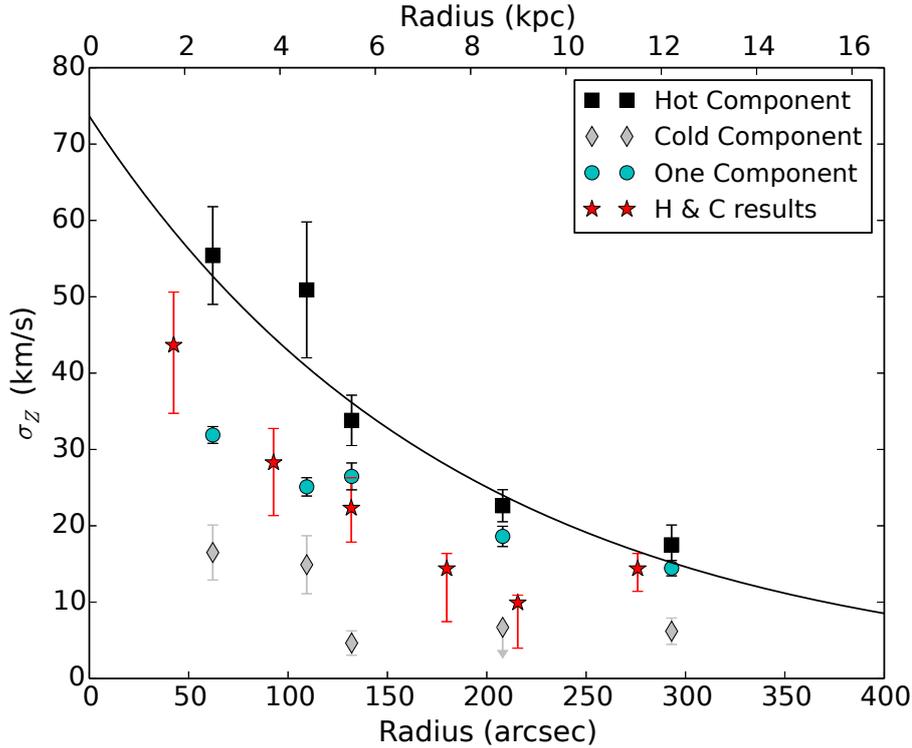

**Figure 10.** The $\sigma_z$ per radial bin in NGC 628. The black and grey markers indicate the hot and cold velocity dispersions respectively from our two component fits, the cyan markers are our single-component values, and the points in red are the Herrmann & Ciardullo (2009b) PNe data. The data points at $R = 62''$ and $109''$ were obtained for integrated light spectra from VIRUS-W and the outer data points are for PNe from PN.S. The solid line denotes an exponential with twice the galaxy's dynamical scale length, fit to the hot component. Our data have been corrected for the HI velocity dispersion and PNe velocity errors (see section 4.2.3 for more details). The errors bars are the $1\sigma$ errors obtained from monte carlo simulations.

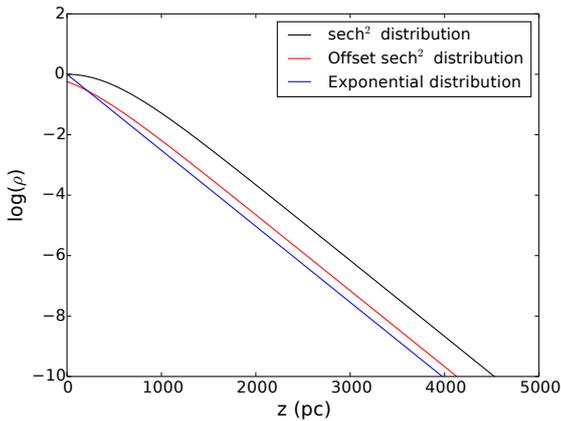

**Figure 11.** An illustration of a sech$^2$ distribution (in red) with an offset parameter of 0.5, taking into account the contribution from the cold layer of gas and young stars. It is identical to an exponential profile (shown in blue) beyond $z \sim 200$ pc. A sech$^2$ distribution without any offset is shown in black for comparison.

From equation A10 in the appendix, the velocity dispersion of the isothermal sheet is $\sigma_z^2 = 2\pi G h_z (\Sigma_C + \Sigma_D)$. The adopted scale height $h_z = 398 \pm 88$ pc (see section 4). If $h_z$ is constant with radius, the radial variation of $\sigma_z^2$ will follow the total surface density rather than the surface density of the isothermal sheet. In NGC 628, the gas fraction of the total surface density is significant at larger radii (Table 6).

### 9.1.1 The surface density of the cold stellar sheet

Our goal is to calculate the rotation curve contribution from the baryons in the disc. Assuming that the disc is a hot isothermal stellar layer $\Sigma_D$ plus a thin young cold layer $\Sigma_C$, we measure the total surface density $\Sigma_T = \Sigma_C + \Sigma_D$. The cold layer $\Sigma_C$ is made up partly of the gas sheet and partly of the thin young stellar disc. The cold stellar layer is thin, and the hot stellar layer is thicker, and their shape affects the calculated rotation curve. This is a second order effect, but it would be useful to know both of the stellar surface densities $\Sigma_D$ and $\Sigma_{C,*}$.

The velocity dispersion analysis gives the total surface density

$$\Sigma_T = \Sigma_D + \Sigma_{C,\text{gas}} + \Sigma_{C,*}$$

of the disc. We can make an estimate of the ratio of cold to hot stellar contributions $\Sigma_{C,*}/\Sigma_D$ from our IFU study in the inner parts of NGC 628 which gives the fraction of light that comes from the two stellar components. Their $M/L$ ratios need to be estimated from stellar population synthesis. In the outer regions, the PNe analysis gives the fraction of cold emission-line objects, but deriving the fractional mass of the underlying cold stellar population is uncertain because (a) of possible contamination of the sample by HII regions, and (b) the young PNe have more massive progenitors, and the lifetimes in the PNe phase are very strongly dependent on their progenitor masses (e.g. Miller Bertolami 2016).

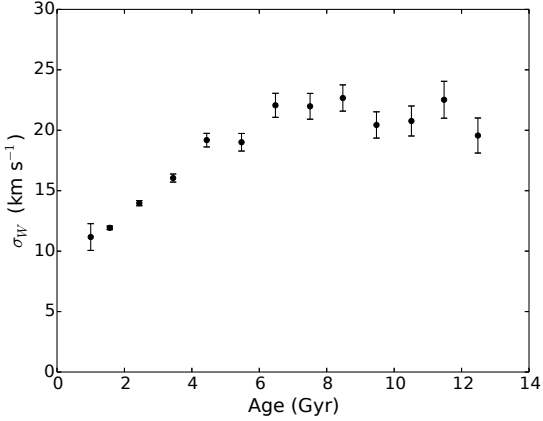

**Figure 12.** Vertical velocity dispersion $\sigma_W$ vs stellar age for the solar neighbourhood, for stars with [Fe/H] > −0.3; data from Casagrande et al. (2011).

Our IFU study analysed the integrated light spectrum as the sum of two velocity components (hot and cold, or young and old), without specifying the age range of either component. The age range is needed to calculate their $M/L$ ratios, and we estimate the age ranges from the observed age-velocity dispersion relation (AVR) in the solar neighbourhood. Figure 12 shows the vertical velocity dispersion $\sigma_W$ vs age for the solar neighbourhood Casagrande et al. (2011) stars from the Geneva Copenhagen survey. In order to exclude most thick disc stars, Figure 12 includes only stars with [Fe/H] > −0.3. The dispersion rises from about 10 km s$^{-1}$ for ages < 1 Gyr to about 20 km s$^{-1}$ at 4 Gyr and then remains approximately constant. If the AVR in NGC 628 has a similar shape, then the cold stellar component would be made up of stars younger than about 3 Gyr.

The (cold, hot) component contributions to the integrated light spectrum shown in Figure 7 (close to the V-band) are given as percentages $L_C/L_D$ in the first two rows of Table 6. From Bruzual & Charlot (2003), we compute the expected $(M/L)_V$ ratios for populations younger and older than 3 Gyr, assuming a disc age of 12 Gyr and a star formation rate that decays with time like $\exp(-t/\tau)$. For $\tau > 3$ Gyr, the $M/L$ ratios for the young and old populations vary with $\tau$ almost in lockstep, and the ratio $F_C = \Sigma_{C,*}/\Sigma_D$ is about 0.12, almost independent of $\tau$. The surface densities of the stellar disc components are then

$$\Sigma_D = \frac{\Sigma_T - \Sigma_{C,\text{gas}}}{1 + F_C} \quad \text{and} \quad \Sigma_{C,*} = F_C \Sigma_D. \quad (8)$$

In the first two rows of Table 6, we use values of $F_C$ derived from the (cold, hot) component contributions to the integrated light spectrum as described above. In the following rows, where the velocity dispersions come from PNe, we adopt the mean value of 0.12 for $F_C$ from the integrated light spectra. Table 6 gives $F_C$, $\Sigma_D$, $\Sigma_{C,*}$ and finally the $(M/L)_V$ ratio for the stellar component at each radius i.e $(\Sigma_D + \Sigma_{c,*})/L$. We can compare the observed values of $(M/L)_V$ with those calculated from the Bruzual & Charlot (2003) models. The calculated $(M/L)_V$ is in the range 2.3 to 2.9 for the star formation rate parameter $\tau > 3$

Gyr, in fair agreement with our values in Table 6 for the stellar disc of NGC 628. The weighted mean of our $(M/L)_{3.6}$ values from Table 6 is 0.75 ± 0.18. This value is similar to the disc stellar $(M/L)_{3.6} \sim 0.6$ obtained from population synthesis studies (Meidt et al. 2014; Norris et al. 2014).

## 10 ROTATION CURVE DECOMPOSITION

### 10.1 Observed HI rotation curve

Obtaining the rotation curve of a near-face-on galaxy, such as NGC 628, is challenging. The uncertainty associated with the inclination is substantial and is the main contributor to the error budget of the rotational velocity. Therefore, we approach this task in several steps. First, we estimate an approximate value for the average inclination angle by placing this galaxy on the 3.6 $\mu$m Tully-Fisher relation from Ponomareva et al. (2017). This gives us an initial guess of the inclination angle = 9°.

We then use the HI data for NGC 628 from the THINGS survey (Walter et al. 2008) to get the rotation curve for this galaxy. After smoothing and masking the original data cube, we constructed a velocity field by fitting a Gauss-Hermite polynomial to the velocity profiles. This velocity field is corrected for the skewness of the profiles, in order to remove non-circular motions, using the technique described in Ponomareva et al. (2016). We then derive the rotation curve from the velocity field by fitting a tilted-ring model (Begeman 1989), using the Groningen Imaging Processing System (Gipsy, Van der Hulst et al. 1992). During the tilted ring modelling, we account for the outer warp of the disc, which can be clearly seen in the upper panel in Figure 13 (see Mulcahy et al. 2017). We find the average inclination across all radii to be 8.5° ± 0.2° (Figure 13, middle panel). We derive the rotation curve of the galaxy by keeping the inclination fixed at this value. The resulting rotation curve is presented in the bottom panel of Figure 13, with red and blue curves indicating the receding and approaching sides of the galaxy respectively. We adopt the difference between the receding and approaching sides as the $1\sigma$ error in the measured rotational velocity. The full error in the rotational velocity was calculated by also taking into account the uncertainty in the inclination.

We now proceed to calculate the contribution to the observed rotation curve from the gas. An integrated column–density HI map was created by summing the primary beam-corrected channels of the clean data cube. The radial surface density profile is then derived by averaging the pixel values in the concentric ellipses projected on to the HI map. We use the same radial sampling, position and inclination as those for the rotation curve derivation.

The column density profile was then corrected for the presence of helium and metals by multiplying the measured densities by 1.4. The Gipsy task ROTMOD was used to compute the corresponding rotation curve assuming an infinitely thin exponential gas disc. In addition to the atomic gas, we also adopt the CO data from Leroy et al. (2009) to derive the rotational velocity for the molecular gas (see section 6 for details).

### 10.2 Stellar Distribution

One of the largest sources of uncertainty for the mass modelling is the stellar disc mass, which usually depends on the adopted

| Instrument | R (kpc) | Colour B-I | $\sigma_z$ (km s$^{-1}$) | $\Sigma_T$ (M$_\odot$ pc$^{-2}$) | $\Sigma_{C,gas}$ (M$_\odot$ pc$^{-2}$) | b | $L_C/L_D$ | $F_C$ | $\Sigma_D$ (M$_\odot$ pc$^{-2}$) | $\Sigma_{C,*}$ (M$_\odot$ pc$^{-2}$) | $(M/L)_B$ | $(M/L)_V$ | $(M/L)_R$ | $(M/L)_I$ | $(M/L)_{3.6}$ |
|---|---|---|---|---|---|---|---|---|---|---|---|---|---|---|---|
| (1) | (2) | (3) | (4) | (5) | (6) | (7) | (8) | (9) | (10) | (11) | (12) | (13) | (14) | (15) | (16) |
| VIRUS-W | 2.6 | 1.82 | 55.4 ± 6.4 | 286 ± 92 | 34 ± 3 | 0.22 ± 0.09 | 44/56 | 0.13 | 223 ± 82 | 29 ± 11 | 2.5 ± 0.9 | 2.3 ± 0.9 | 2.3 ± 0.9 | 1.9 ± 0.7 | 0.8 ± 0.3 |
| VIRUS-W | 4.5 | 1.70 | 50.9 ± 8.9 | 241 ± 100 | 34 ± 3 | 0.23 ± 0.11 | 40/60 | 0.11 | 187 ± 90 | 21 ± 10 | 3.5 ± 1.7 | 3.3 ± 1.6 | 3.5 ± 1.7 | 2.9 ± 1.4 | 1.2 ± 0.6 |
| PN.S | 5.5 | 1.64 | 33.8 ± 3.3 | 106 ± 31 | 36 ± 3 | 0.44 ± 0.16 | 42/58 | 0.12 | 63 ± 28 | 8 ± 3 | 1.5 ± 0.7 | 1.5 ± 0.7 | 1.6 ± 0.7 | 1.4 ± 0.6 | 0.6 ± 0.3 |
| PN.S | 8.7 | 1.44 | 22.6 ± 2.1 | 48 ± 14 | 25 ± 3 | 0.68 ± 0.29 | 42/58 | 0.12 | 21 ± 13 | 3 ± 2 | 1.2 ± 0.7 | 1.2 ± 0.8 | 1.4 ± 0.9 | 1.3 ± 0.8 | 0.6 ± 0.4 |
| PN.S | 12.2 | 1.22 | 17.5 ± 2.6 | 29 ± 11 | 14 ± 2 | 0.64 ± 0.32 | 42/58 | 0.12 | 13 ± 10 | 2 ± 1 | 2.0 ± 1.5 | 2.3 ± 1.7 | 2.8 ± 2.2 | 2.6 ± 2.0 | 1.3 ± 0.9 |

**Table 6.** Parameters for NGC 628 at the five radii where the velocity dispersion of the hot disc was measured (inner two radii from integrated light, outer three from PNe). All surface densities are in units of M$_\odot$ pc$^{-2}$. The columns are (1) instruments used, (2) radius, (3) B - I colour from Möllenhoff (2004) (4) vertical velocity dispersion $\sigma_z$ of the hot disc, (5) total surface density $\Sigma_T$ from equation A10 in the appendix, (6) observed surface density $\Sigma_{C,gas}$ of the gas layer from the THINGS & HERACLES survey, (7) the offset parameter $b$ from equation A5 in the appendix, (8) ratio of luminosities of cold and hot layers from the integrated spectra in rows 1 and 2, and adopting the mean of rows 1 and 2 in rows 3 to 5, (9) the ratio $F_C$ of the stellar surface densities of the cold and hot layers, (10) the surface density $\Sigma_D$ of the hot layer from equations 8, (11) the stellar surface density $\Sigma_{C,*}$ of the cold stellar layer, (12) – (16) the foreground extinction corrected $M/L$ ratio of the total stellar component in BVRI bands from Möllenhoff (2004) and the 3.6 $\mu$m band from Salo et al. (2015).

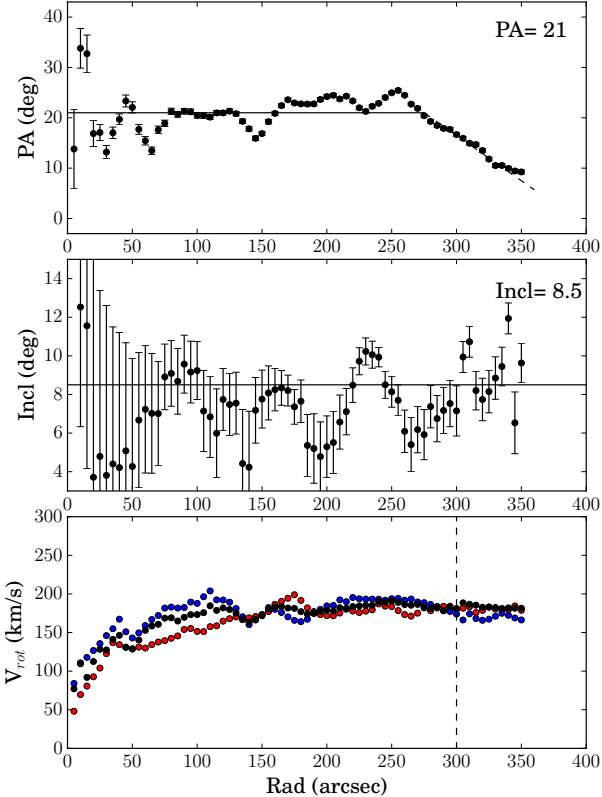

**Figure 13.** The results of the tilted-ring modelling for NGC 628. Upper panel: position angle as a function of radius. The outer warp is modelled beyond 250″. Middle panel: inclination angle as a function of radius. The mean value of 8° is adopted for our analysis. Bottom panel: observed HI rotation curve is shown in black. The red and blue curves show the velocity of the receding and approaching sides of the galaxy respectively. The dashed vertical line indicates the optical radius.

*M/L*. However, for NGC 628 we have measured the surface density of the stellar component directly. We measured the dynamical scale length for the total surface mass density $\Sigma_T$ (see section 8). This included the cold and hot stellar disc as well as the gas disc. However, we do not have a handle on the scale height of the total baryonic disc component represented by $\Sigma_T$. The adopted scale height of 398 pc is for the old disc component. The scale height of $\Sigma_T$ will be some combination of the scale heights of the cold and hot component. In order to test the effect of various scale heights on the rotation curve decomposition, we adopted a range of scale heights from 100 – 400 pc. Figure 14 shows the effect of the scale heights on the rotation curve of the baryonic component.

As discussed in section 4, a small bulge component is present in NGC 628 (Muñoz-Mateos et al. 2013). While the total bulge light contributes only 6.5% to the total light of the galaxy, the bulge light dominates within the central 1.5 kpc and, therefore, we need to include it in our mass modelling. We convert the 3.6 $\mu$m bulge surface brightness from Salo et al. (2015) into the surface mass density using the stellar *M/L* for 3.6 $\mu$m = 0.6 (Meidt et al. 2012; Querejeta et al. 2015; Röck et al. 2015) and $M_\odot$(3.6 $\mu$m) = 3.24 mag (Oh et al. 2008). We model the bulge using the same Gipsy task ROTMOD assuming a spherical potential. The rotation curves of all the baryonic compo-nents were modelled with the same sampling as was used for the derivation of the observed rotation curve.

Figure 14 shows the observed rotation curve in black. The dashed curves in various colours are the bulge + the central total surface mass density with a dynamical scale length ($h_{dyn}$ = 92.7″) fit as an exponential disc with varying scale heights. We note that a difference of 100 pc in scale height only changes the peak of the rotational velocity of the baryonic component ($V_{max}$) by 1 km s$^{-1}$. Thus the adopted scale height has a negligible effect on the rotation curve decomposition.

### 10.3 Mass modelling

The total rotational velocity of a spiral galaxy can be calculated using the relation:

$$V_{tot}^2 = V_{bar}^2 + V_{halo}^2, \quad (9)$$

where $V_{bar}$ is the circular velocity associated with the gravitational field of the total baryonic content: cold disc+hot disc+gas, and $V_{halo}$ is the circular velocity for the dark halo. As the observed HI rotation curve ($V_{obs}$) traces the total gravitational potential of a galaxy, we can derive the rotation curve of the dark matter halo by fitting various parameters until $V_{tot}$ matches $V_{obs}$ as closely as possible. Moreover, for this galaxy, we have an independently measured surface density distribution for the stellar disc components, so we do not have the usual unconstrained stellar *M/L*. This allows us to fix the baryonic rotation curve and only fit the rotation curve of the dark matter halo. This breaks the usual disc-halo degeneracy and allows us to test the maximality of the disc directly.

For our analysis we use two models of the dark matter rotation curves: spherical pISO (pseudo isothermal) and NFW-halo. The pISO halo has a central core and is parameterised by its central density ($\rho_0$) and the radius of the core ($R_c$). Its rotation curve is given by:

$$V_{DM}^{pISO}(R) = \sqrt{4\pi G \rho_0 R_C^2 \left[1 - \frac{R_C}{R} tan^{-1}\left(\frac{R}{R_C}\right)\right]} \quad (10)$$

The NFW halo (Navarro et al. 1997) is parameterised by its mass ($M_{200}$) within the viral radius ($R_{200}$) and its concentration ($c$). Its rotation curve is given by:

$$V_{DM}^{NFW}(R) = V_{200}\left[\frac{ln(1+cx) - cx/(1+cx)}{x[ln(1+c) - c/(1+c)]}\right]^{1/2} \quad (11)$$

where $x = R/R_{200}$ and $V_{200}$ is the circular velocity at $R_{200}$.

Equations 10 and 11 describe the circular velocity distribution of the dark halo as it is now, modelled by the two particular models which we are using (pISO and NFW). The state of the dark halo now is almost certainly very different from what it was like before the baryons condensed to form the disc. Based on the present understanding of the formation of the dark halo, it probably formed with a density cusp as seen in almost all N-body simulations (e.g. Navarro et al. 2010). Cusped haloes are however rarely observed in disc galaxies at low redshift (e.g. Oh et al. 2011). Their absence is usually interpreted as the effect of feedback from rapid star formation as the stellar disc is forming (e.g. Brook et al. 2011), which transforms the inner cusps into the approximately flat cores that are observed.

In addition to this major restructuring of the dark haloes, many authors have considered the effect of slow (adiabatic) formation of the disc and bulge on the large scale structure of the dark halo. Blumenthal et al. (1986) proposed a simple method for calculating the restructuring of the dark halo under adiabatic contraction of the disc and bulge, based on the adiabatic invariance of the angular momentum of dark matter particles. Subsequent works have argued that this simple method overestimates the readjustment of the halo relative to calculations in which the other adiabatic invariants are also included (Wilson 2003; Gnedin et al. 2004; Jesseit et al. 2002). Abadi et al. (2010) use N-body simulations and find that the halo contraction is less pronounced than found in earlier simulations, a disagreement which suggests that halo contraction is not solely a function of the initial and final distribution of baryons. Wegg et al. (2016) compared the dark halo rotation curves of the Milky Way obtained from the MOA-II microlensing data (see their Figure 12). They find that the adiabatically contracted halo according to the classical Blumenthal et al. (1986) prescription is inconsistent with the revised MOA-II data at the 1$\sigma$ level. The prescription of Gnedin et al. (2004) is consistent only if the initial halo has low concentrations (see Wegg et al. 2016 for details).

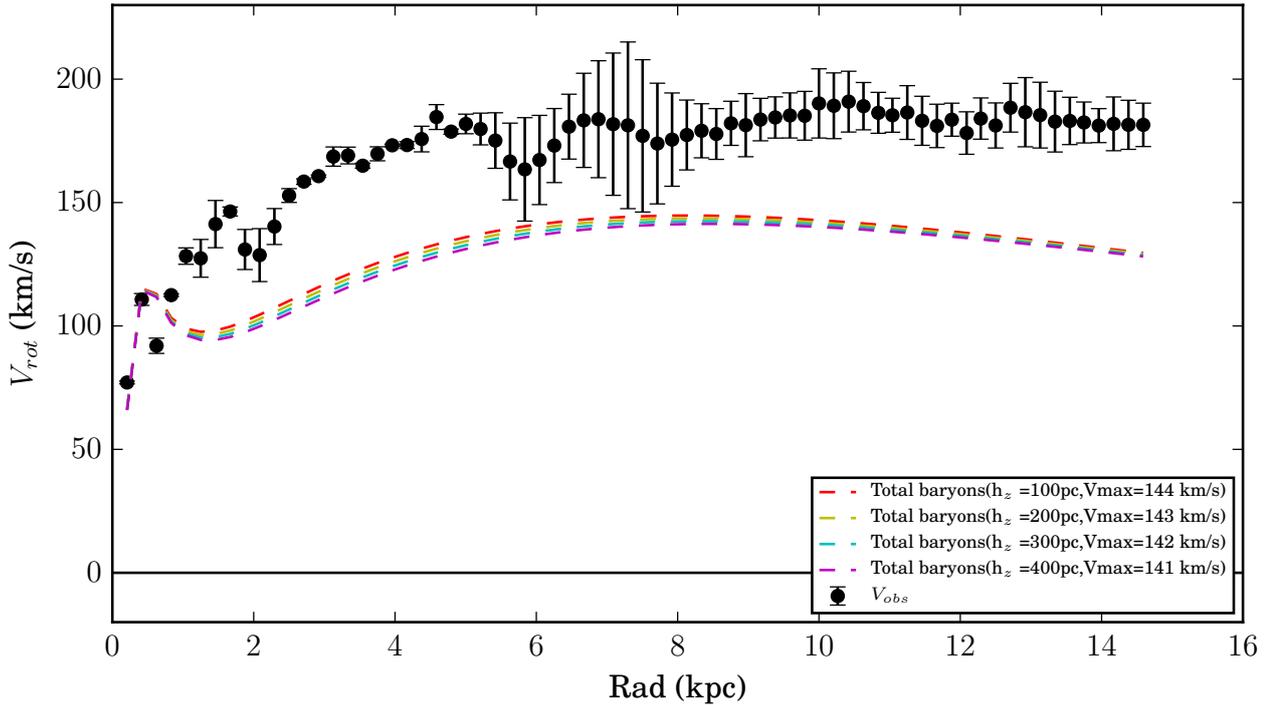

**Figure 14.** The effect of the scale height on the rotation curve decomposition. The black data points are the observed HI rotation curve from the THINGS survey. The coloured dashed lines are the total baryonic rotation curve assuming scale heights of 100 – 400 pc. A difference of 100 pc in the scale height results in a 1 km s$^{-1}$ change in the peak of the baryonic rotation.

The evolution of dark haloes is not simple. We have briefly discussed cusp-core transformation, which probably occurs during the strong star formation around $z = 2$ to 3. Adiabatic contraction, if it is significant, is presumably an extended process that goes on throughout the growth of the disc. Furthermore, cosmological simulations indicate that haloes continue to grow by accretion up to the present time, with typically half or more of their mass growth taking place since $z = 1$ (Muñoz-Cuartas et al. 2011). While we can measure the basic parameters of dark haloes at the present time, as described in this paper, we do not see an obvious earlier stage at which one would want to identify a pristine dark halo and estimate its parameters from its present state. We do not therefore attempt to go beyond our initial goal, of determining the decomposition of the rotation curve of galaxies into the contributions from the disc and dark halo at the present time.

The Gipsy task ROTMAS was again used to find the rotation curves for the dark haloes. We ran the models for each of the two dark matter haloes separately, keeping the rotation curves of the total baryonic component fixed. As demonstrated in the previous section, the scale height makes a negligible contribution to the rotation curve decomposition. We therefore fit an exponential with $\Sigma_T(0) = 505 \pm 175$ M$_\odot$pc$^{-2}$ and (dynamical) scale length = 92.7″ along with a scale height = 397.6 pc (this is our adopted scale height for the hot stellar disc). The results from this modelling are presented in Figure 15 (the top panel is for the pISO halo and the bottom panel for the NFW halo). Although the rotation curves for the two haloes have a similar shape, the pISO halo works marginally better in our case with a $\chi^2_{red} = 1.17$ compared to a $\chi^2_{red} = 1.21$ for the NFW halo.

There is a significant negative covariance between the $h_{dyn}$ and the $\sigma_z(0)$. Taking our $\sigma_z(0)$ in units of km s$^{-1}$ and $h_{dyn}$ in units of arcsecs, the cov($\sigma_z(0), h_{dyn}$) = $-118.12$. The fractional error on the peak of the baryonic rotation curve can be calculated as:

$$\left(\frac{\Delta V_{baryonic}}{V_{baryonic}}\right)^2 = \left(\frac{\Delta\sigma_z(0)}{\sigma_z(0)}\right)^2 + 0.25\left(\frac{\Delta h_{dyn}}{h_{dyn}}\right)^2 + 0.25\left(\frac{\Delta h_z}{h_z}\right)^2 + \frac{\text{cov}(\sigma_z(0), h_{dyn})}{\sigma_z(0).h_{dyn}}$$

where the $\Delta$ terms represent the errors on the respective values. The corresponding values for our data are:

$$\left(\frac{\Delta V_{\text{baryonic}}}{V_{\text{baryonic}}}\right)^2 = 0.018 + 0.005 + 0.012 - 0.017$$

This gives us a 13% error on the peak of the baryonic rotation curve. Therefore, our estimated $V_{\text{baryonic}} = 141.0 \pm 18.8$ km s$^{-1}$ and $V_{max} = 180 \pm 9$ km s$^{-1}$ (from our observed rotation curve). Taking into account the covariance between the terms, we find our disc to be maximal with $V_{\text{baryonic}} = (0.78 \pm 0.11)V_{max}$. We note that the $M/L$ for the bulge component remains at 0.6 even when the bulge rotation curve was not constrained to this value. The derived parameters of the fitted dark matter rotation curves are presented in Table 7. We note that the scale length derived for the pISO model is rather short for a luminous galaxy such as NGC 628 (Kormendy & Freeman 2016). The error on the total rotational velocity (blue curve in Figure 15) follows from the error on the central surface density described in section 5.

| pISO | | | NFW | | |
|---|---|---|---|---|---|
| $R_C$ (kpc) | $\rho_0$ ($10^{-3}$ M$_\odot$pc$^{-3}$) | $\chi^2_{red}$ | C | $R_{200}$ (kpc) | $\chi^2_{red}$ |
| 0.69 ± 0.09 | 642.9 ± 159.8 | 1.17 | 25.1 ± 2.1 | 117.2 ± 4.1 | 1.21 |

**Table 7.** The derived parameters of the fitted dark matter haloes from the mass modelling.

The derived dark halo parameters for the pseudo isothermal and NFW models have significant covariance. This is demonstrated using the $\chi^2$ maps of the 2D parameter space as shown in Figure 16. The coloured ellipses in each panel represent the $1\sigma$ to $5\sigma$ values from inside to outside. The errors quoted in Table 7 are the $1\sigma$ values.

For the purpose of the rotation curve analysis shown in Figure 15, we simply needed the radial distribution of the total baryon surface density $\Sigma_T$. It may however be of interest for the reader to see how each of the components of $\Sigma_T$ contributes to the rotation curve. We are able to derive the radial surface density distributions separately for the hot and cold stellar discs, as described in section 9.1.1, although extra errors are introduced in the process. The surface density distribution of the gas is observed directly.

The dynamical scale length derived from Figure 10 refers to the total baryonic surface density of the disc of NGC 628. To demonstrate the contribution of the individual components to the rotation curve, we need to estimate their scale lengths. Figure 17 shows our $\Sigma_D$ and $\Sigma_{C,*}$ radial distributions from Table 6. Fitting exponential profiles to these distributions gives an estimate of the scale lengths for the hot and cold stellar components as $\sim 65'' \pm 17''$. The errors are relatively large, as we have only five points on each profile. We note that the two stellar components are expected to have equal scale lengths, from the way in which they were derived; see equation 8.

Figure 18 shows the contribution of the various baryonic components to the rotation curve. We stress that this figure is for the purpose of illustration only, and that our adopted parameters for the individual stellar components have large errors, as explained above. In Figure 18 the green dot dashed line represents the rotational velocity of the atomic + molecular gas, using data from the THINGS and HERACLES surveys (Walter et al. 2008; Leroy et al. 2009). The red dot-dashed line is the bulge component. The blue and red dashed line represent the cold and hot stellar disc respectively. The purple curve with error bars is the total baryonic rotational velocity. This is a bit different from the corresponding curve in Figure 15 due to the significant errors involved in deriving the parameters of the hot and cold stellar discs individually.

## 11 CONCLUSIONS

The disc-(dark halo) degeneracy in spiral galaxies has been an important problem in dark halo studies for several decades. One of the more direct methods of breaking this degeneracy is by measuring the surface mass density of the disc and hence its mass-to-light ratio, using the velocity dispersion and the estimated scale height of the disc. However, it is essential that the measured vertical velocity dispersion and the disc scale height pertain to the same stellar population.

In the disc of a typical star-forming spiral galaxy, the colder young stellar component provides a significant contribution to the number density of kinematical tracers like red giants and planetary nebulae which have stellar progenitors with a wide range of ages. Through its red giants, which are a significant contributor to the integrated light spectra that are used for velocity dispersion measurements, the colder young population can have a substantial influence on the observed vertical velocity dispersion of the disc.

The other observable, the scale height of the disc, is usually measured from red or near-IR surface photometry of edge-on galaxies, away from the central dust layer. The estimated scale heights are then close to the scale heights of the older disc population.

Using integrated light absorption line spectra and velocities for a large sample of PNe in the face-on disc galaxy NGC 628, we show that the velocity distribution of disc stars does indeed include a young kinematically cold population and an older kinematically hotter component. We argue that previous integrated light and PNe studies have underestimated the surface density and $M/L$ ratio of discs because they were not able to account for the contribution from the younger colder disc population in their velocity dispersion analysis.

To estimate the surface density of the disc, using equation 2, we use the velocity dispersion $\sigma_z$ of the older hotter component of the disc together with the old disc scale height $h_z$ measured from red or near-IR photometry of edge-on spiral galaxies. Together, $\sigma_z$ and $h_z$ pertain to the same stellar population and act as consistent tracers of the total gravitational field of the baryonic disc (hot + cold stellar disc + gas disc). For NGC 628, this leads to a surface mass density that is $\sim 2$ times larger than the values derived from previous estimates of disc density from disc velocity dispersions. This factor agrees well with our earlier estimates for the solar neighbourhood (Aniyan et al. 2016), and is large enough to make the difference between concluding that a disc is maximal or sub-maximal.

We find that the observed vertical velocity dispersion of the hotter component follows an exponential radial decrease, as expected for an exponential disc of constant scale height. From the velocity dispersion distribution, we estimate a dynamical scale length and central total surface density for the disc, which are then used in the rotation curve modelling. In the presence of a thin cold layer of gas and young stars with surface density $\Sigma_C$, the density distribution of the hot isothermal layer has the form

$$\rho(z) = \frac{\sigma_z^2}{8\pi G h_z^2} \text{sech}^2(|z|/2h_z + b)$$

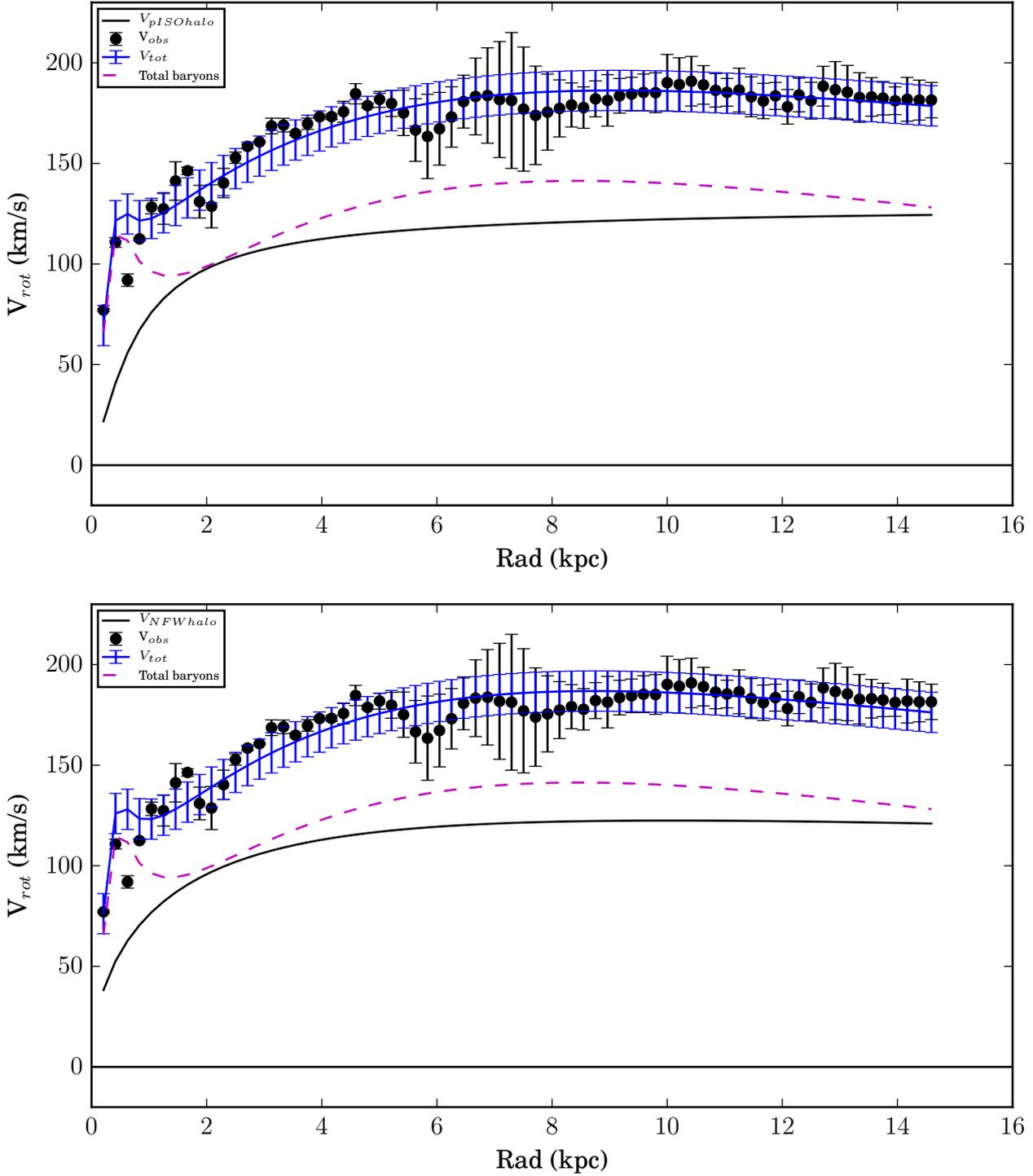

**Figure 15.** The rotation curve decomposition for NGC 628 by fitting a pISO halo (top panel) and a NFW halo (bottom panel). Observed HI rotation curve is shown as black dots. The magenta line represents the rotational velocity of all the baryons i.e the disc, gas, and the bulge. The black curve shows the rotation of the dark halo. This galaxy is clearly maximal demonstrated by the fact that the baryons are dominant in the inner parts of the galaxy.

where $\tanh(b) = 2\pi G h_z \Sigma_C / \sigma_z^2$. The total surface density is then $\Sigma_T = \Sigma_C + \Sigma_D = \sigma_z^2 / 2\pi G h_z^2$. We find that the ratio $F_c = \Sigma_{C,*}/\Sigma_D$ appears to be about 12% in the inner regions of of NGC 628, and our calculated M/L for the total stellar component (hot + cold) is approximately constant with radius.

Decomposing the rotation curve of this galaxy, after taking into account the hot and cold stellar components, leads to a maximal disc. The rotation of the baryonic component is ∼ 78% of the total rotational velocity of this galaxy.

This is the first report from a larger study of nearby near-face-on spirals, using a combination of VIRUS-W data and PN.S data for 5 northern galaxies and data from the WiFeS IFU spectrograph for 3 southern galaxies. The analysis of these galaxies is currently underway.

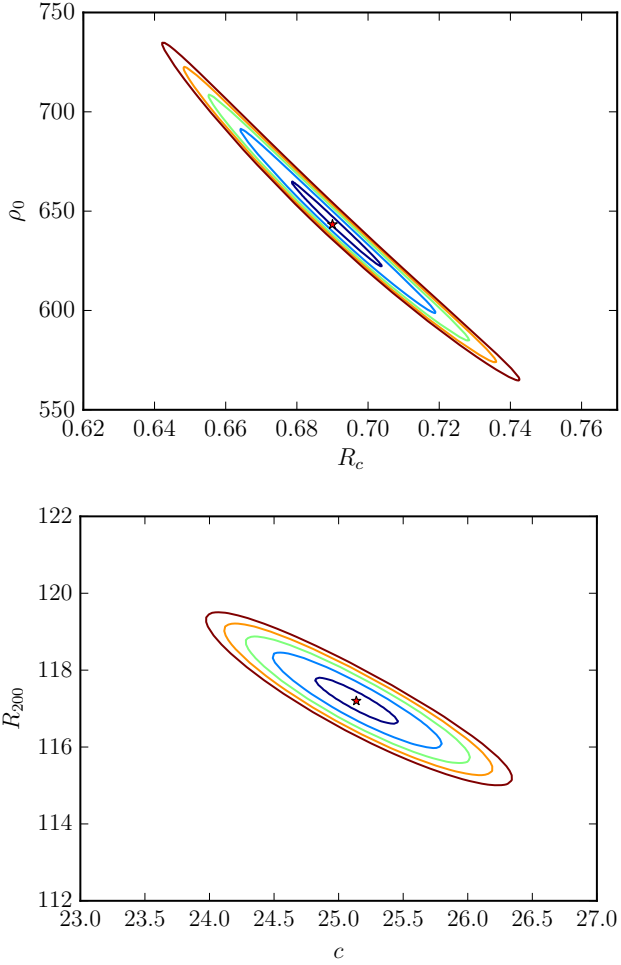

**Figure 16.** $\chi^2$ map in the two dimensional parameter space for our derived dark halo parameters. The upper panel shows the covariance between the density and radius of the core of the pseudo isothermal model of the dark halo. The lower panel shows the covariance between the virial radius and concentration of the NFW dark halo model. The ellipses represent the $1\sigma - 5\sigma$ values from inside out.

## 12 ACKNOWLEDGEMENTS

The authors would like to thank the referee for the very detailed report that improved the quality of this paper. SA would like to thank ESO for the ESO studentship that helped support part of this work. KCF thanks Roberto Saglia, Marc Verheijen and Matt Bershady for helpful discussions on this project. KCF, MA, and OG acknowledge the support of the ARC Discovery Project grant DP150104129. The authors are very grateful to the staff at McDonald Observatory for granting us the observing time and support on the 107″ telescope. The authors would also like to thank the Isaac Newton Group staff on La Palma for supporting the PN.S over the years, and the Swiss National Science Foundation, the Kapteyn Institute, the University of Nottingham, and INAF for the construction and deployment of the H$\alpha$ arm.

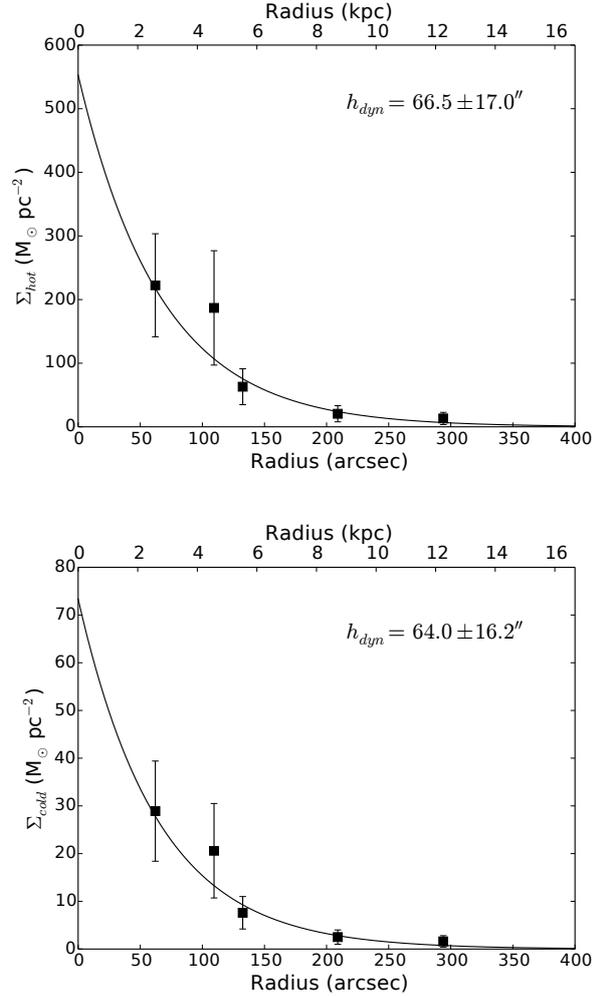

**Figure 17.** The surface mass density of our hot stellar component (upper panel) and cold stellar component (lower panel) to derive the dynamical scale length.

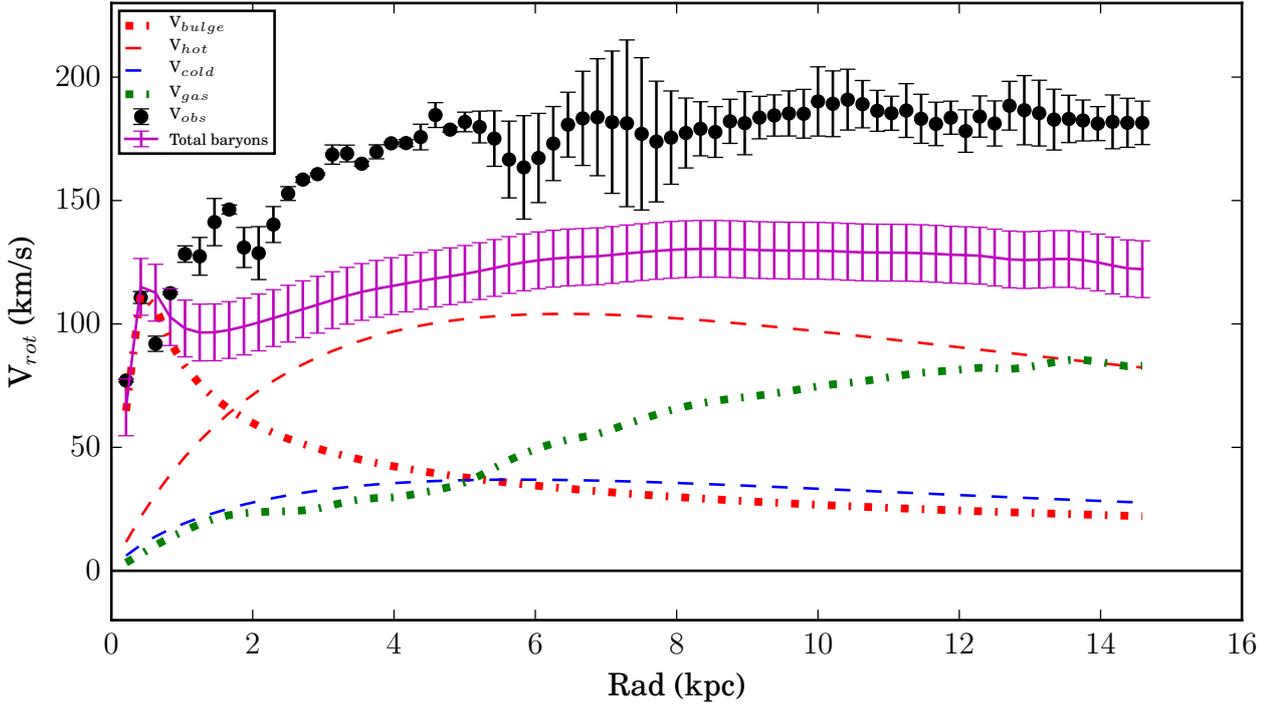

**Figure 18.** A demonstration of the contributions of the various baryonic components to the rotation velocity. The green dot dashed line represents the rotational velocity of the atomic + molecular gas, using data from the THINGS and HERACLES survey. The red dot dashed line is the bulge component. The blue and red dashed line represent the cold and hot stellar disc respectively. The purple curve is the total baryonic rotational velocity obtained by summing up each component in quadrature.

# APPENDIX A: EFFECT OF THE COLD LAYER ON THE EQUILIBRUM OF THE ISOTHERMAL SHEET

Previous studies that looked at calculating the surface mass density of the disc, ignored the gravitational field of the cold layer (gas and young stars). In this Appendix, we estimate the effect of the cold layer on the structure and surface density of the isothermal disc (see Binney & Tremaine II: problems 4.21 & 4.22).

## A1 The distribution function for the isothermal disc

The stellar energy is $E = v^2/2 + \Phi$, where $v$ is the stellar velocity and $\Phi$ the potential. The distribution function has the form $f(E) \propto \exp(-E/\sigma^2)$ where $\sigma$ is the stellar velocity dispersion. Write $\Phi = \sigma^2 \phi$. The distribution function for the isothermal sheet is

$$f = \frac{\rho_\circ}{\sqrt{2\pi\sigma^2}} \exp\left(-v^2/2\sigma^2 - \phi\right).$$

The velocity distribution is isothermal and everywhere gaussian, and the density is

$$\rho(z) = \rho_\circ \exp(-\phi)$$

where we take $\phi(0) = 0$. Write $z/h_z = \zeta$. The 1D Poisson equation is

$$\frac{d^2\Phi}{dz^2} = 4\pi G \rho$$

or

$$2\frac{d^2\phi}{d\zeta^2} = \frac{8\pi G \rho_\circ h_z^2}{\sigma^2} \exp(-\phi)$$

or

$$2\frac{d^2\phi}{d\zeta^2} = \exp(-\phi) \quad (A1)$$

for $h_z^2 = \sigma_z^2/(8\pi G \rho_\circ)$. Equation A1 has the solution $\phi = \ln \cosh^2(\zeta/2)$, $d\phi/d\zeta = \tanh(\zeta/2)$, and

$$\rho = \rho_\circ \exp(-\phi) = \rho_\circ \operatorname{sech}^2(\zeta/2) = \rho_\circ \operatorname{sech}^2(z/2h_z).$$

At large $|z|$, $\rho \to \exp(-|z|/h_z)$. The surface density is $\Sigma = 4\rho_\circ h_z$, and the vertical force is $K_z = -8\pi G \rho_\circ h_z \tanh(z/2h_z)$.

## A2 The isothermal sheet in the presence of the cold layer

We model the disc as an isothermal sheet of hot disc population, plus a thin layer of cold disc population (gas and young stars) with surface density $\Sigma_C$. The total density is then

$$\rho_D(z) + \Sigma_C \delta(z)$$

where $\rho_D(z)$ is the density distribution of the isothermal sheet in the presence of the extra component. The vertical force and potential for the cold layer are:

$$K_z = -2\pi G \Sigma_C z/|z|$$
$$\Phi_C = 2\pi G \Sigma_C . |z|$$

With the extra component, the energy is $E = v^2/2 + \Phi$, where $\Phi = \Phi_D + \Phi_C$ and $\Phi_D$ is the potential of the hot isothermal sheet. Write $\Phi = \sigma_z^2 \phi$. The distribution function for the isothermal sheet is

$$f_D = \frac{\rho_\circ}{\sqrt{2\pi\sigma^2}} \exp\left(-v^2/2\sigma_z^2 - \phi\right)$$

and

$$\rho_D = \rho_\circ \exp(-\phi). \quad (A2)$$

Write $z/h_z = \zeta$. The 1D Poisson equation for the isothermal sheet is

$$2\frac{d^2\phi_D}{d\zeta^2} = \frac{8\pi G \rho_\circ h_z^2}{\sigma_z^2} \exp(-\phi_D - \phi_C)$$

where $\phi_C = 2\pi G \Sigma_C h_z . |\zeta|/\sigma_z^2$. With $\phi = \phi_D + \phi_C$, equation A2 becomes

$$2\frac{d^2\phi}{d\zeta^2} = \frac{8\pi G \rho_\circ h_z^2}{\sigma_z^2} \exp(-\phi) \quad (A3)$$

for $|\zeta| > 0$. The solution to equation A3 gives the potential $\phi$ and the density distribution $\rho_D$ for the isothermal sheet in the presence of the cold layer. We look for a solution of the form

$$\phi = \ln \frac{\cosh^2(|\zeta|/2 + b)}{\cosh^2 b} \quad (A4)$$

such that $\phi(0) = 0$ and $\phi' = \tanh(|\zeta|/2 + b) \to 2\pi G h_z \Sigma_C/\sigma_z^2$ as $\zeta \to 0$. Equation A4 is a solution if

$$\tanh(b) = 2\pi G h_z \Sigma_C/\sigma^2 \quad (A5)$$

and

$$8\pi G \rho_\circ \cosh^2(b) h_z^2/\sigma_z^2 = 1. \quad (A6)$$

The density of the isothermal sheet is then

$$\rho_D(\zeta) = \rho_\circ \cosh^2(b) \operatorname{sech}^2(|\zeta|/2 + b). \quad (A7)$$

and its surface density is

$$\Sigma_D = \frac{4\rho_\circ h_z}{1 + \tanh(b)}. \tag{A8}$$

### A2.1 The surface density of the isothermal disc

At large $|z|$, equation A7 becomes

$$\rho_D \propto \exp(-|z|/h_z)$$

and the photometric scale height is approximately equal to the natural scale parameter $h_z$ defined in equation A6. From equations A6 – A8, we eliminate the parameter $\rho_\circ$. In terms of its surface density $\Sigma_D$ and the observables $\Sigma_C$, $h_z$ and $\sigma_z$, the density distribution of the isothermal sheet is

$$\rho(z) = \frac{\sigma_z^2}{8\pi G h_z^2} \text{sech}^2(|\zeta|/2 + b), \tag{A9}$$

the scaling equation A6 becomes

$$\Sigma_T = \Sigma_C + \Sigma_D = \frac{\sigma_z^2}{2\pi G h_z}, \tag{A10}$$

and equation A5 defines the offset parameter $b$. From equations A5 and A10, $b$ is approximately the surface density fraction $\Sigma_C/(\Sigma_C + \Sigma_D)$ in the cold layer.

### A3 Summary

- In the presence of the cold gas layer $\Sigma_C$, the isothermal sheet has the form

$$\rho(z) = \frac{\sigma_z^2}{8\pi G h_z^2} \text{sech}^2(|z|/2h_z + b)$$

where $\tanh(b) = 2\pi G h_z \Sigma_C / \sigma_z^2$.
- The total surface density $\Sigma_T = \Sigma_C + \Sigma_D = \sigma_z^2/(2\pi G h_z)$.
- The scale parameter $h_z$ for the isothermal sheet $\approx$ the exponential scale height for the hot disc.
- The radial dependence of $\sigma_z^2(R) \propto \Sigma_T(R)$, if $h_z$ is constant.